\documentstyle{mn}
\begin{document}
\def\simlt{\mathrel{\rlap{\lower 3pt\hbox{$\sim$}}
        \raise 2.0pt\hbox{$<$}}}
\def\simgt{\mathrel{\rlap{\lower 3pt\hbox{$\sim$}}
        \raise 2.0pt\hbox{$>$}}}

\title[$N(z)$ for FIRST Radio Sources]
{The Redshift Distribution of FIRST Radio Sources at 1~mJy}
\author[M. Magliocchetti, S.J.Maddox, J.V.Wall, C.R.Benn, G.Cotter]
{M.~Magliocchetti $^{1,2}$,
S.J.Maddox ${^3}$, J.V.Wall $^4$  C.R.Benn$^5$, G.Cotter$^6$\\
$^1$Institute of Astronomy, Madingley Road, Cambridge CB3 0HA\\
$^2$SISSA Via Beirut 4, 34100, Trieste Italy \\
$^3$School of Physics and Astronomy, University of Nottingham,
Nottingham NG7 2RD, UK \\
$^4$Astrophysics, Nuclear and Astrophysics Laboratory, Keble Road,
Oxford, OX1 3RH, UK \\
$^5$ Isaac Newton Group, Apartado 321, 38700 Santa Cruz de La Palma,
The Canary Islands, Spain \\
$^6$ Cavendish Laboratory,  Madingley Road,  Cambridge CB3 0HE, UK 
}

\maketitle
\vspace {7cm }

\begin{abstract}
We present spectra for a sample of radio sources from the FIRST
survey, and use them to define the form of the redshift distribution
of radio sources at mJy levels.
We targeted 365 sources and obtained 46 redshifts (13 per cent of the
sample). 
We find that our sample is complete in redshift measurement to R $\sim
18.6$, corresponding to $z\sim 0.2$. Galaxies were assigned spectral 
types based on emission line strengths. 
Early-type galaxies represent the largest subset (45 per cent) of the
sample and have redshifts $0.15\la z \la 0.5$ ; late-type galaxies
make up 15 per cent of the sample and have redshifts $0.05\la z \la
0.2$; starbursting galaxies are a small fraction ($\sim 6$ per cent),
and are very nearby ($z\la 0.05$).
Some 9 per cent of the population have Seyfert1/quasar-type
spectra, all at $z\ga 0.8$, and there are 4 per cent are Seyfert2 type galaxies
at intermediate redshifts ($z\sim 0.2$). 


Using our measurements and data from the Phoenix survey (Hopkins et al.,
1998), we obtain an estimate for $N(z)$ at $S_{1.4 \rm {GHz}}\ge 1$
mJy and compare this with model predictions. At variance with previous
conclusions, we find that the population of starbursting objects makes
up $\la 5 $ per cent of the radio population at S $\sim 1$~mJy.

\end{abstract}


\section{INTRODUCTION}

An accurate definition of the redshift distribution of radio sources
at faint flux densities has become particularly important in the last
decade for both radio astronomy and cosmology. It is critical, for
example, in testing radio-source unification (e.g. Jackson \& Wall,
1998), and in large-scale structure studies (e.g. Loan et al. 1997,
Magliocchetti et al. 1999) to permit the conversion of angular
clustering estimates to the spatial clustering estimates required to
evaluate structure formation models.
 
Classical large-scale-structure studies have used wide-area optical
and IR surveys to measure the clustering of galaxies, but these are
limited to low redshifts, with a peak redshift selection of $\la 0.1$
(e.g. APM, Maddox et al., 1990; IRAS, Fisher et al., 1993). Deep
small-area surveys such as the Hubble Deep Field North (Williams et
al., 1996) or the CFRS survey (Lilly et al., 1995) have been used to
measure clustering to much higher redshifts (see e.g. Le Fevre et al.,
1995; Magliocchetti \& Maddox, 1998), but they probe small volumes and
can measure clustering only on small scales ($\la 1$ Mpc). On the
other hand radio objects are detected to high redshifts
($z\sim 4$) and sample a much larger volume for a given number of
galaxies, so that they have the potential to provide information on
the growth of structure on large physical scales. Recently, clustering
statistics in deep radio surveys have been measured (Cress et al.,
1997; Loan et al., 1997; Baleisis et al., 1997; Magliocchetti et al.,
1998) and these analyses have shown radio sources to be reliable
tracers of the mass distribution, even though they may be biased
(Magliochetti et al., 1999).  

However the relationship between angular measurements and the
physically meaningful spatial quantities is highly uncertain  
without an accurate estimate of the redshift distribution of radio objects 
$N(z)$ (Magliocchetti et al.,
1999). The current estimates of $N(z)$ for radio sources at mJy levels
are largely based on predictions from the local radio
luminosity functions and evolution (see e.g. Dunlop \& Peacock, 1990), and are poorly defined.
 
Indeed the predictions require the knowledge of more than one
luminosity function. Deep radio surveys (Condon \& Mitchell, 1984;
Windhorst et al., 1985; Fomalont et al., 1993) have shown a flattening
of the differential source count $\frac{dN}{dS}$ below S $\sim 10$
mJy. This flattening is generally interpreted as due to the presence
of a population of radio sources differing from the radio AGN which
dominate at higher flux densities. Condon (1984) suggested a
population of strongly-evolving normal spiral galaxies, while others
(Windhorst et al., 1985; Danese et al., 1987) claimed the presence of
an actively star-forming galaxy
population. Observations supported this latter suggestion; the
identifications of many of the faint sources are with galaxies with
spectroscopic and photometric properties similar to `IRAS
galaxies' (Franceschini et al., 1988; Benn et al., 1993). The model
predictions of $N(z)$ at faint flux densities from the luminosity
functions of Dunlop \& Peacock (1990) diverge greatly because of
inadequate definition of the radio AGN luminosity functions; but in
addition the contribution to $N(z)$ of the starburst population needs
consideration.  

In order to define $N(z)$ and the population mix at mJy levels, we
have observed radio sources from the FIRST survey (Becker et al.,
1995) using the WYFFOS multi-object spectrograph on the William
Herschel Telescope in 8 1-degree diameter fields. 
We show how these observations constrain both the form of
$N(z)$ and the population mix at S $\sim 1$~mJy for $z\la 0.3$. Section
2 of the paper introduces this radio sample, while Sections 3 and 4
describe acquisition and reduction of the data. In Sections 5, 6 and 7
we present the radio, spectroscopic and photometric properties of the
sample. Section 8 is devoted to the analysis of the observed $N(z)$
and comparison with models; the conclusions are in Section 9.
\begin{figure} 
\vspace{19cm}  
\includegraphics{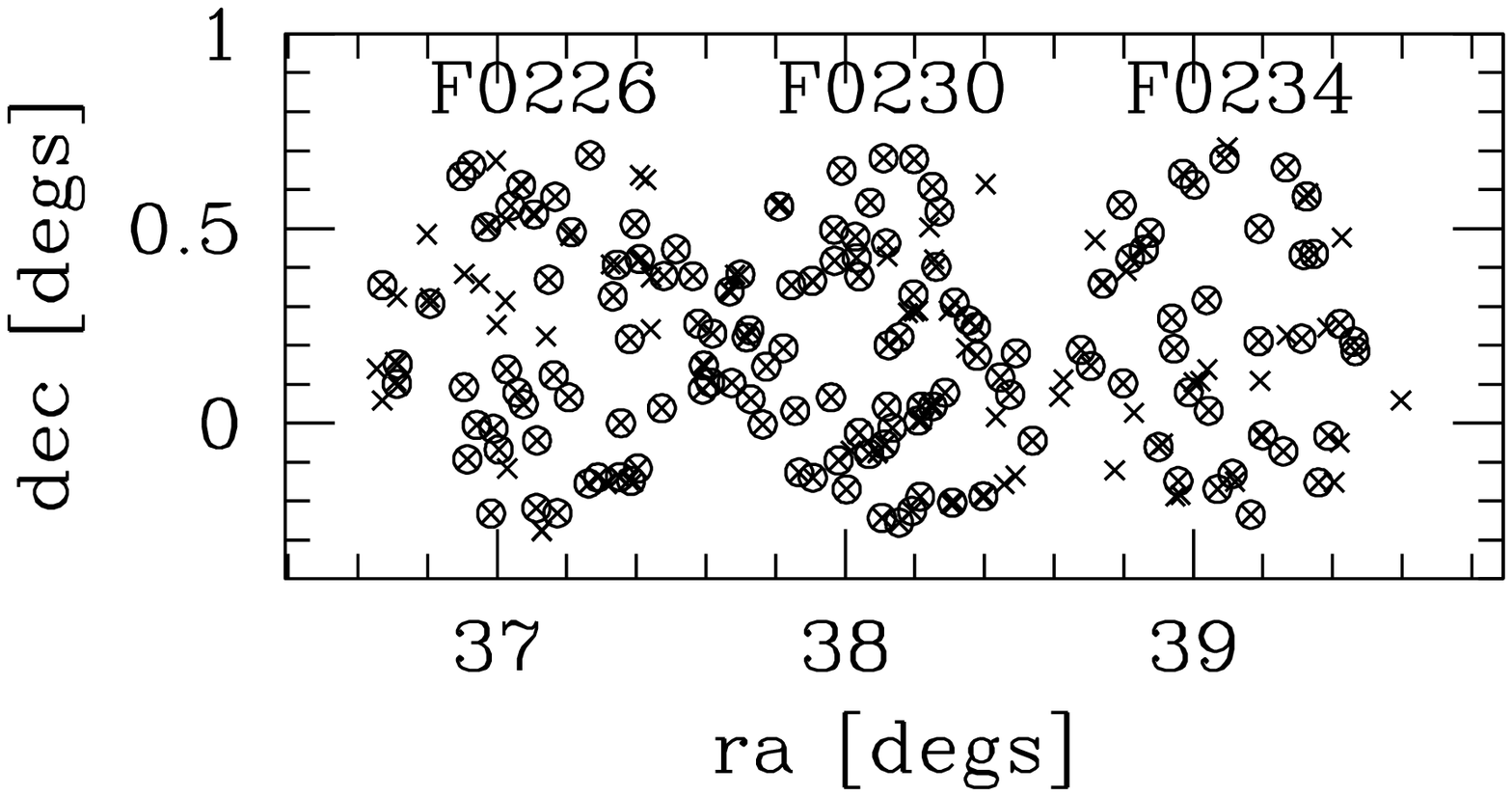}
\includegraphics{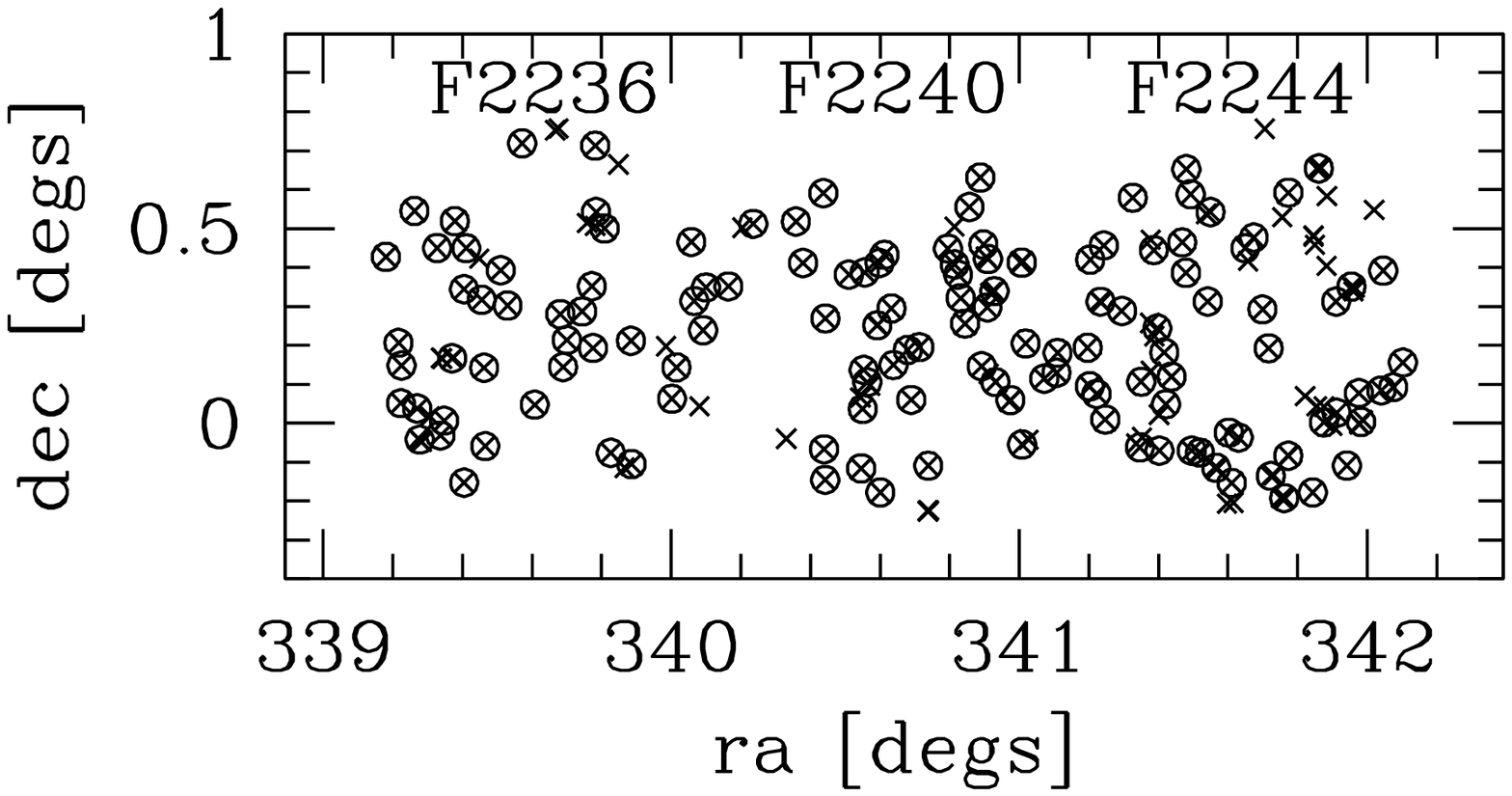}
\includegraphics{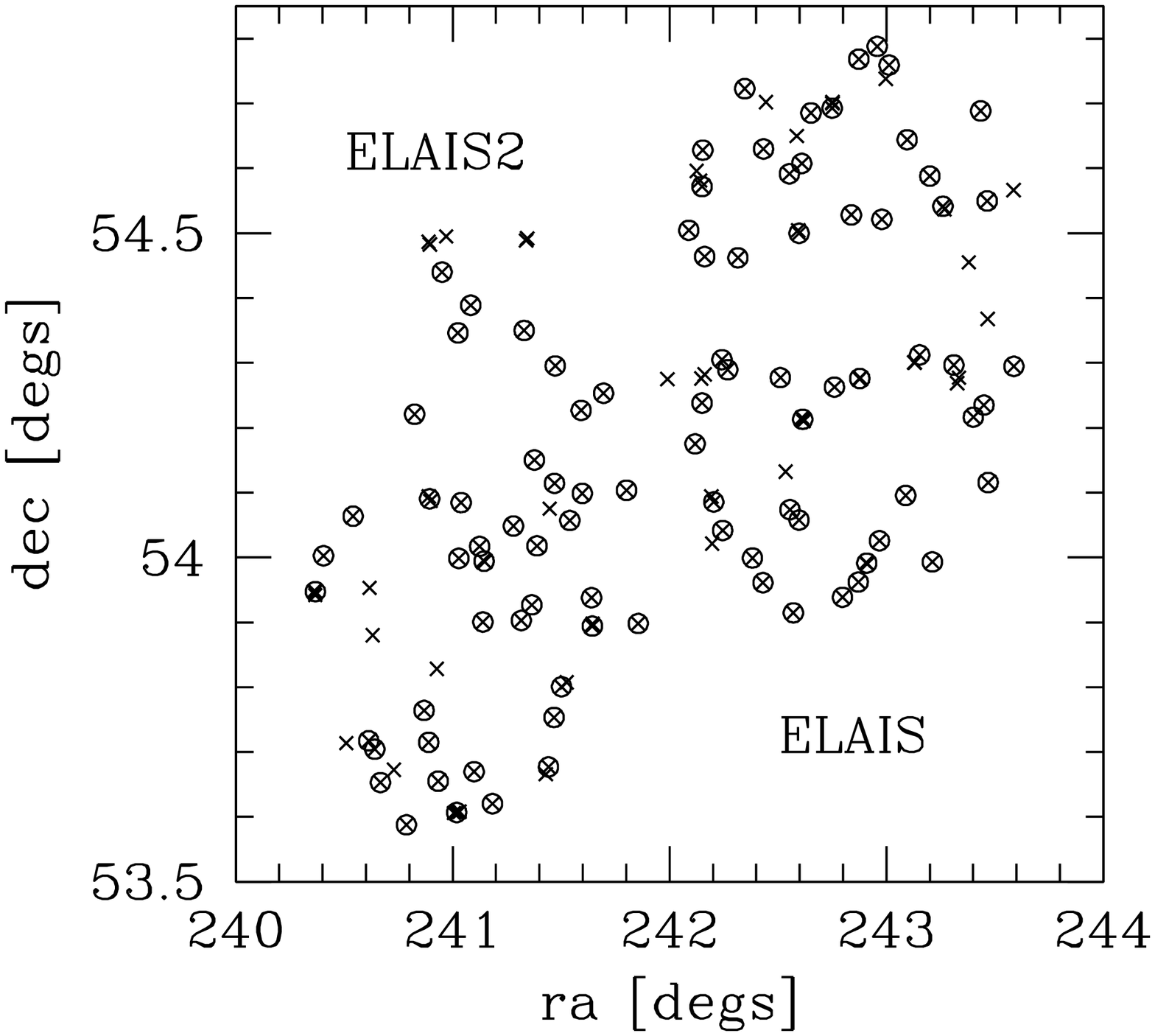}
\caption{Area covered by the 8 observed fields. Crosses are for
  the radio sources while open circles show the
  positions of the optical fibres. 
\label{fig:fields}} 
\end{figure}

\section{THE RADIO DATA}

The FIRST (Faint Images of the Radio Sky at Twenty centimetres) survey
(Becker, White and Helfand, 1995) began in the spring of 1993 and will
eventually cover some 10,000 square degrees of the sky in the north
Galactic cap and equatorial zones.  The VLA is being used in
B-configuration to take 3-min snapshots of 23.5-arcmin fields arranged
on a hexagonal grid.  The beam-size is 5.4~arcsec at 1.4~GHz, with an
rms sensitivity of typically 0.14~mJy/beam.  A map is produced for
each field and sources are detected using an elliptical Gaussian
fitting procedure (White et al., 1997); the $5\sigma$ source detection
limit is $\sim$1~mJy.  The surface density of objects in the catalogue
is $\sim 90$ per square degree, though this is reduced to $\sim 80$
per square degree if we combine multi-component sources 
(Magliocchetti et al., 1998).
The depth, uniformity and angular extent of the survey are excellent
attributes for investigating the clustering properties of faint
sources.

We used the 4 April 1998 version of the catalogue which contains
approximately 437,000 sources and is derived from the 1993 through
1997 observations covering nearly 5000 square degrees of sky,
including most of the area $7^h20^m < {\rm RA}(2000) < 17^h20^m$,
$22.2^\circ < {\rm Dec} < 57.5^\circ$ and $21^h20^m < {\rm RA}(2000) <
3^h20^m$, $-2.5^\circ < {\rm Dec} < 1.6^\circ$. This catalogue has been
estimated to be 95 per cent complete at 2~mJy and 80 per cent complete
at 1~mJy (Becker et al.,~1995).
  
\section {OBSERVATIONS}

\subsection{ Field Selection} 
For the purpose of determining the redshift distribution, the precise
position of the fields is not crucial, although several factors were
taken into account when choosing which to observe. 
The fields must span a range of RA's to ensure they could be
observed at low zenith distance throughout the allocated nights. 
The fields are at as high galactic latitude as possible to reduce
the probability of confused  optical identifications and to ensure low
galactic extinction (the allocated time meant the fields had to be on
opposite sides of the galactic plane). 
We also cross-checked with fields which will be observed as part of the 
INT Wide-Field Camera survey, and chose our fields to overlap with
these, so that deep optical images will be available in the near future.

These constraints lead to the choice of three main areas: $16^h11^m, \;
+54^\circ 22'$ (which covers the ISO ELAIS field); $22^h40^m, \; +0^\circ 0'$,
and $02^h30^m, \; +0^\circ 0 '$. We also observed a field adjacent to the
ELAIS field, and one on each side of the $22^h40^m, \; +0^\circ 0'$, and 
 $02^h30, \; +0^\circ 0 '$ fields. The areas actually observed are shown
in Figure \ref{fig:fields}, and the field centres are listed in
Table~\ref{Table_fields}. 

\subsection {WYFFOS and AUTOFIB2} 
The 4.2-m William Herschel Telescope (WHT) has a prime-focus corrector
with an atmospheric-dispersion compensator that provides a field of
view of one degree in diameter.  AUTOFIB2 is a robotic positioner,
which places up to 120 optical fibre feeds in the focal plane, each
fibre collecting light from a $2.7''$ diameter patch of sky.  The
fibres are fed to WYFFOS which is a multi-object spectrograph, at one
of the WHT Nasmyth foci.  The spectrograph can use a range of gratings
offering 30 - 500 \AA/mm dispersion. We chose to use the 600 lines/mm
grating with the Tek1024$^2$ CCD, which gives a dispersion of $\sim 3$
\AA\ per pixel, and a resolution of $\sim 10$\AA.  For a central
wavelength at 6000\AA\ this set up gives spectra from $\sim 4400$\AA\
to $\sim 7400$\AA.

\subsection{ Astrometry and Fibre Positioning} 

The fibres on WYFOSS cover 2.7 arcsecs on the sky, but to obtain the
best throughput, they must be positioned to an accuracy of $0.5''$ or
better.  The astrometric reference frame of the FIRST survey is accurate
to $0.05''$, and individual sources have 90\% confidence error circles of
radius $< 0.5''$ at the 3 mJy level and $1''$ at the survey threshold. This
means that, in principle, the radio source positions are accurate enough
to position the fibres, even without optically identified counterparts. 
Using the radio positions means that our sample is not biased towards
optically bright sources. Although our spectroscopy will clearly be
limited by the optical brightness, it should be possible to obtain
redshifts for any sources that have strong optical emission lines but
very faint continuum. These sources would have been rejected from the
sample if we required every target to have an optical counterpart. 
However there are a few complications in using the radio positions and
these must be taken into account.

First, the telescope must be positioned and guided using fiducial stars.
We selected stars with $13 < b_J < 15 $ from APM scans of UKST plates
for the equatorial fields, and of POSS2 plates for the ELAIS fields.
These star positions are accurate to about $0.1''$, but have been
measured in an optical reference frame, which  may be offset from the
radio reference frame by more than an arcsecond.  Therefore we matched
up the radio sources and optical images for each Schmidt plate.

\begin{figure} 
\vspace{10cm}  
\caption{The positional residuals between FIRST radio sources and APM
optical positions in the ELAIS field. The points with residuals
greater than $1''-2''$ are likely to be random coincidences and 
the points with near-zero offset are the true optical
identifications.  
The scatter suggests that the combined rms error
between the optical and radio objects is about $0.5''$, which is
consistent with the quoted errors in each survey.  The mean offset is
$0.9''$ in RA and $0.5''$ in Dec. 
\label{fig_astrom} 
}
\includegraphics{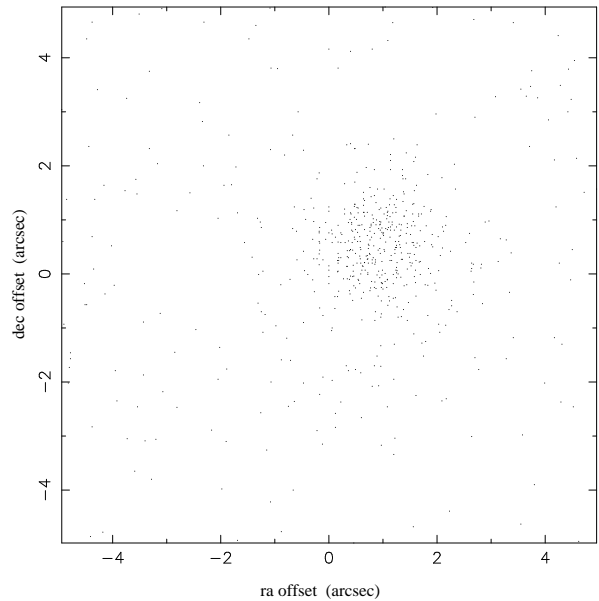}
\end{figure}

Figure~\ref{fig_astrom} shows a typical plot of the positional
residuals between the optical and radio positions.  There is a uniform
background of points with residuals greater than $1''-2''$, which come
from random coincidences between optical and radio sources. The true
optical identifications show up as the concentration of points near
zero offset.
The mean positional offset of the identifications is $0.9''$ in RA and
$0.5''$ in Dec, corresponding to the offset between the optical and
radio astrometric frames, and this offset was applied to each position
in the ELAIS field so that the optical and radio astrometry 
coincide.  Similar offsets were calculated for each of the other
observed fields.
The scatter in residuals about the mean offset is about $0.5''$ which
is consistent with the quoted positional accuracy of the surveys.

\begin{figure}
\vspace{10cm}  
\caption{A grey-scale $4'\times4'$ radio map of a FIRST source. 
The overlaid contour map is from the optical DSS image. 
\label{fig_radio_optmap} 
}
\includegraphics{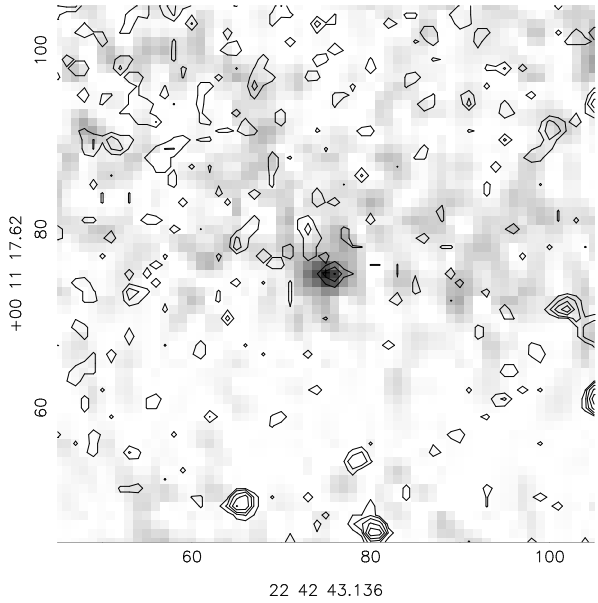}
\end{figure}

The second complication is that the
optical ID of an extended radiosource will not necessarily lie at the
radio catalogue position. In the case of a classical double-lobed
radiosource, the radio core and optical ID will lie between the lobes,
which themselves may be catalogued as two separate radio sources. In
the case of a core-jet radiosource, the radio core and optical ID will
usually lie away from the centroid of the radio emission.
To address this problem, we extracted $4'\times4'$ 
radio maps from the FIRST database and optical images covering a similar
area from the Digital Sky Survey (DSS). We visually inspected each of 
the radio maps, and overlaid the optical contours so that we could
identify possible offsets between the radio and optical emission. 
Almost all of the radiosources were either unresolved or
dominated by an unresolved component in the FIRST maps.
Only in a
few cases it was necessary to manually adjust the positions. 
A typical plot showing the FIRST radio map and overlaid optical contours
is shown in Figure~\ref{fig_radio_optmap}. 

As well as the fiducial stars and target objects, it is necessary to
observe some regions of blank sky, so that the sky spectrum during the
observations can be estimated and subtracted from the target spectra. 
On average there are 70 radio sources per field (see
Table~\ref{Table_fields}), 
leaving 50 fibres which
could be used for sky. 
Two sets of sky positions were generated for each field. The first set
were the target positions in the focal plane when the telescope is 
offset by 5 arcminutes in Dec. Note that the spherical geometry of the
sky means that new positions do not exactly correspond to a simple
addition of 5 arcmins to each Dec.  
The low surface density of bright optical images meant that these
positions were blank sky for all but a couple of cases. 
The second set were simply randomly chosen positions; again most of these
are blank sky. 
Given the large number of sky fibres we could median them together to
obtain a highly accurate estimate of the true sky spectrum, even if a
few were contaminated by objects. 

\subsection{Field Configurations} 

\begin{figure} 
\vspace{10cm}  
\caption{The fibre configuration for a FIRST field. The large circle
represents the full - 1 degree - field of view. The dark grey outlines
show the positions of fibres for the spectrograph and the black
outlines show the fibres for the fiducial stars. Broken or unusable
fibres are shown in light grey.  The FIRST sources are marked with crosses,
and the sky positions are marked by circles.  
\label{fig_configure} 
}
\includegraphics{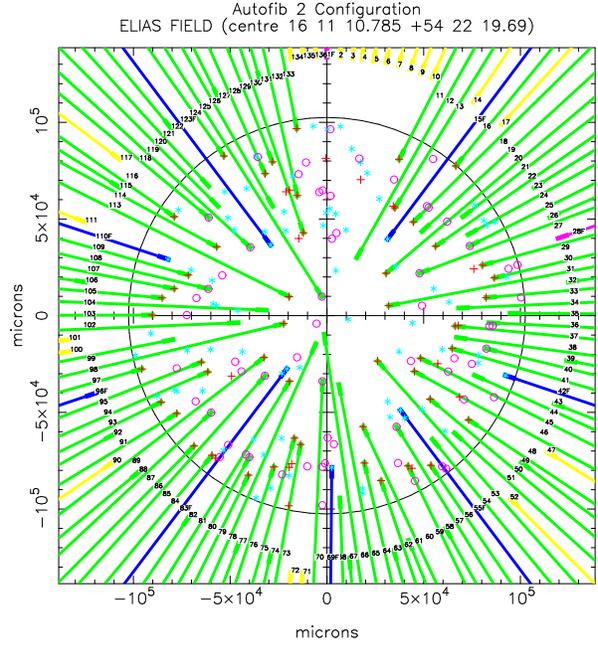}
\end{figure}

Each fibre for AUTOFIB2 can be placed within a restricted area defined by
a maximum radial extension and a maximum tilt angle from the radial
direction. Also no fibre can cross another, and  there must be a small
buffer-zone around each fibre.  These constraints limit the number of
targets that can be allocated to a fibre within a particular field.  The
program CONFIGURE is part of the WYFFOS software package, and it
attempts to find the most efficient set of target-to-fibre allocations. 
This configuration was then inspected and some changes were made 
manually to increase the number of fibres allocated to targets. 
A typical configuration is shown in Figure~\ref{fig_configure}.  
The total number of targets and number of fibres allocated is shown in
Table~\ref{Table_fields}.

\begin{center}
\begin{table*}
\caption{Observed fields. $\alpha$ and $\delta$ are respectively the
  RA and Dec of the field centres. $N_{\rm radio}$ is the number of 
radio objects in each field, $N_{\rm obs}$ is the number of
object-positions on which fibres
could be placed and $N_z$ is the number of sources with a redshift
determination. Note that $N_{\rm radio}$ is the number obtained after
combining multiple components into a single source. \label{Table_fields} } 
\begin{tabular}{llllllllll} 
& & & & & & &\\
  Field & $\alpha$ (J2000)& $\delta$ (J2000)  & $N_{\rm obs}$ & $N_{\rm
radio}$ & \% Configured & $N_z$ & \% Redshifts \\
& & & & & & &\\
\hline
{\rm F0226} & 02 28 33.8 & +00 13 22 & 48 & 78& 61.5 & 3 & 6.3 \\
{\rm F0230} & 02 32 33.8 & +00 13 12 & 57 & 86& 66.3 & 4 & 7.0 \\
{\rm F0234} & 02 34 33.8 & +00 13 01 & 35 & 55& 63.6 & 1 & 2.8 \\
{\rm F2236} & 22 38 33.7 & +00 15 38 & 40 & 53& 75.5 & 10 & 25 \\ 
{\rm F2240} & 22 42 33.7 & +00 15 44 & 44 & 59& 74.5 & 10 & 23 \\
{\rm F2244} & 22 46 33.7 & +00 15 49 & 50 & 85& 58.8 & 11 & 22 \\
{\rm ELAIS} & 16 10 00.0 & +54 30 00 & 49 & 72& 68.0 & 3 & 6.1 \\
{\rm ELAIS2}& 10 04 13.0 & +54 01 52 & 42 & 62& 67.7 & 4 & 9.5  \\
\end{tabular} 
\end{table*}
\end{center}

\subsection{Observing Procedure} 

Once a field has been configured and acquired we took a standard sequence of
calibration frames and observations of the target field. 
The sequence began with an arc lamp for wavelength calibration; then
1800 sec on the targets;  then three 600 sec offset sky integrations; 
another 1800 sec on the targets; another three 600 sec
skies; and finally 1800 sec on targets. Each of the offset skies was 
given a different offset to ensure that the sky spectra were not
contaminated by random objects that happen to fall on the fibres. 

Most of the targets are very faint, so accurate sky subtraction is crucial. 
The main aim of the observing sequence is to provide an
accurate estimate of the sky spectrum during the observations, and also
to calibrate the relative throughputs of each fibre for this particular
configuration. 
Since the field is vignetted, the apparent transmission of a fibre will
depend on its position in the field. Also  the efficiency of light
entering a fibre will depend on the precise angle that the fibre button
settled when placed by AUTOFIB2, so there may be small variations even if
the fibres are repositioned to the same configuration. This also means
that flat-fields from internal lamps will not exactly reproduce the
throughputs as seen on the sky. 
The offset skies are observed with
exactly the same set-up as the target objects, and so provide very
accurate relative throughput estimates for the object and sky fibres. 
The large number of sky fibres provides an accurate estimate of the
actual sky spectrum during the object integrations. 

In principle the $5'$ offset skies allow a more efficient observing mode
that does not require separate offset sky integrations. 
If the telescope is nodded back and forth between the target positions,
and the offset fibres, light from the targets is always being
collected. 
Also the sky  spectrum can be measured through the
actual fibre used for each target. 
The disadvantage of this approach is
that a sky fibre must be allocated to each object fibre, and this leads
to quite severe configuration constraints. Typically less than 50\% of
the available targets could be allocated a pair of fibres, and so this
turned out to be a rather inefficient option for our project. 
Nevertheless we did successfully apply the method for one field (F0234).

During daylight we took high signal-to-noise tungsten lamp
flat-fields which were used to define the position of each fibre
spectrum as described below. 

\section{Data Reduction}

The data obtained were reduced using 
a package
within {\sc  IRAF} written specifically to reduce WYFFOS data. 
The first steps in the data reduction are the standard bias level
subtraction and trimming of the overscan regions.  
Then the tungsten flat-fields are used to locate and fit polynomials
to the spectrum of each fibre. These polynomials allow 1-D spectra to
be extracted from the 2-D images.  The package uses information in the
file headers to ensure the correct link between fibre number and
spectrum number, so that each spectrum is identified with the correct
input target.

Next, for each target field, the 2-D images from the individual
integrations were averaged together, using the {\sc IRAF} ccdreject
option. 
Similarly the offset skies for each field were combined. This way of
combining the images provides good signal-to-noise for pixels where
all component images are good data, and effectively rejects
pixels contaminated by cosmic rays or random objects that happened to
fall in a sky fibre for one particular offset.

As a simple but effective method of sky subtraction, the 2-D averaged
offset skies were rescaled to match the exposure time in the target
frame and then simply subtracted from the 2-D target data.
This technique was used because WYFFOS suffers from light
scattered within the spectrograph.
The broad scattered light features seen in Figure~\ref{fig_2dframe}(a)
have $\sim 10\%$ of the counts in the actual spectra, and so it is
very important to subtract them accurately.
The standard approach attempts to fit a smooth 2-D 
polynomial to the light in between the fibres, and subtracts this from
the 2-D data frame. Then 1-D spectra are extracted from the 2-D data,
and the sky-fibres are used to define the mean sky spectrum, which is
rescaled according to a throughput map generated from offset skies, and
subtracted from each object spectrum. 
We found that it was very difficult to fit the scattered light
accurately enough to leave a clean 2-D image with reliable spectra. 
Since our target galaxies are very faint, the scattered light is
almost entirely from the sky, even in the target fields. Therefore
simply re-scaling the 2-D offset sky frames and subtracting from the
2-D target frame means the scattered light {\em and} the sky are both
properly subtracted. This is very effective, as seen in
Figure~\ref{fig_2dframe}(b).

\begin{figure*} 
\vspace{9cm}  
\caption{(a) A combined 2-D frame of WYFFOS spectra before sky
subtraction. The spectrum from each fibre is dispersed in the vertical
direction, and several bright sky emission lines can be seen in each
spectrum. 
Each set of 3 fibres is offset on the slit in a saw-tooth
configuration, which gives a vertical shift to the wavelength
zero-point of neighbouring spectra. 
The broad horizontal features are
from scattered light in the spectrograph.
(b) the sky-subtracted 2-D spectra. Both the sky lines and the
scattered light have been subtracted well, but the pattern of the
scattered light can be seen as regions of increased noise in the
background. 
Note that the spectra are not perfectly straight, and a polynomial
must be fitted to each one in order to properly extract the 1-D
spectra. 
\label{fig_2dframe} 
}
\end{figure*} 

This sky-subtraction approach is susceptible to errors if the sky is
rapidly varying, because the actual sky spectrum during the target
integration is not used. Any such errors can be simply corrected by
using the spectra from the sky fibres in sky-subtracted 2-D images to
estimate the average residual sky which can then subtracted from the
target spectra.
As can be seen in the spectra, the residuals from sky-subtraction are
very small, confirming that the approach works very well.

The wavelength calibration for each data-set was carried out using the
standard {\sc  IRAF} tasks. 
This automatically accounts for the 'saw-tooth' arrangement of fibres on
the spectrograph slit. 

\section{Spectral Analysis and Classification}

\subsection{REDSHIFT DETERMINATION}

Redshifts were determined in 46 objects, corresponding to $\sim 13$
per cent of the spectroscopic sample. Determinations were obtained in
two separate ways: first by visual inspection of the spectra and
attempting to identify individual features; second
via cross-correlation with a range of templates.

The cross-correlation analysis uses the same algorithm as the 2dF
redshift survey code (Seaborne, 2000) but has small changes to the
filtering parameters, the wavelength coverage and allowed redshift
range appropriate for ou WYFOSS data.
The cross-correlation functions between the data and a set of template
spectra are calculated and the highest peak is located. An index 
$0\le q \le 4$ is automatically
assigned according to the height of the peak
and the noise and assesses the quality of the redshift measurement: 0
corresponding to no redshift estimate, and 4 a reliable redshift. 
Each redshift estimate was also checked by eye at this stage, and the
quality flag was updated if necessary.  
We accepted estimates with $q$ as low as 3,
so long as they agreed with the results from
visual inspection. A few objects (all absorption systems) with $q=2$
were included in the final sample, when the cross-correlation estimate
was in agreement with the results from visual inspection.
Table~\ref{Table_OIDs} lists the measured redshifts for the observed
targets. 

\subsection{SPECTRAL CLASSIFICATION}

The optical counterparts of radio sources with a redshift determination
were classified into 6 main groups according to their
spectral features:
\begin{enumerate}
\item {\it Early-type galaxies} where spectra were dominated by continua
much stronger than the intensity of any emission lines present. These objects
can then be divided into two further categories:
\begin{itemize}
\item Galaxies with absorption lines only.
\item Galaxies with absorption lines + weak but non-negligible OII.
  emission lines
\end{itemize}
\item {\it Late-type galaxies} showing strong emission lines
characteristic of star-forming activity, together with a detectable
continuum.
\item {\it Starburst galaxies} where the continuum is missing and
the spectra only showed strong emission lines due to star-formation
activity.
\item {\it Seyfert1 galaxies} showing strong broad emission
lines. These objects could be quasars.
\item {\it Seyfert2 galaxies} showing narrow emission lines due
to the presence of an active galactic nucleus. To have an objective
classification of the narrow emission line objects and in particular
to distinguish Seyfert2's from galaxies heated by OB stars (classes
2-3), we used the diagnostic emission line ratios of Baldwin,
Phillips \& Terlevich, 1981 and Rola, Terlevich \& Terlevich, 1997.
\item A final class consisted of those objects whose spectra show a
detectable continuum but the complete absence of features; no redshift
determinations for these were possible.  These spectra are generally
believed to be associated with early-type galaxies (see Gruppioni,
Mignoli \& Zamorani, 1998), given the lack of emission lines.
\end{enumerate}
Also, four spectra appear to be stars at zero redshift; these are most
likely to be random positional coincidences with galactic stars.

Table~\ref{Table_OIDs} lists the spectral classes.  All spectra with
significant signal are shown in Figure~\ref{fig:spectra1}.  The
spectrum at the bottom of each panel show the sky spectrum, with
strong emission lines marked. The galaxy spectra have not been flux
calibrated, and the line on the lower right is an approximate
atmospheric transmission curve, showing the position and shape of the
atmospheric A and B absorption bands.

\begin{figure*} 
\vspace{25cm}  
\vspace{-3cm} 
\caption{ Spectra of all detected sources. Each spectrum is labelled
with the object name and redshift where it has been possible to
measure it.  The spectrum at the bottom of each panel is the sky
spectrum, with strong emission lines marked. The galaxy spectra have
not been flux calibrated, and the line on the lower right shows an
approximate atmospheric transmission curve, in particular showing the
position and shape of the atmospheric A and B absorption bands.
\label{fig:spectra1} 
}
\end{figure*}
\begin{figure*} 
\vspace{25cm}  
\vspace{-3cm} 
\contcaption{Spectra of all detected sources}
\label{fig:spectra2} 
\end{figure*}
\begin{figure*} 
\vspace{25cm}  
\vspace{-3cm} 
\contcaption{Spectra of all detected sources}
\label{fig:spectra3} 
\end{figure*}
\noindent
\begin{figure*} 
\vspace{25cm}  
\vspace{-3cm} 
\contcaption{Spectra of all detected sources}
\label{fig:spectra4} 
\end{figure*}

 \section{Photometry}
\begin{figure*}
\vspace{13cm}  
\includegraphics{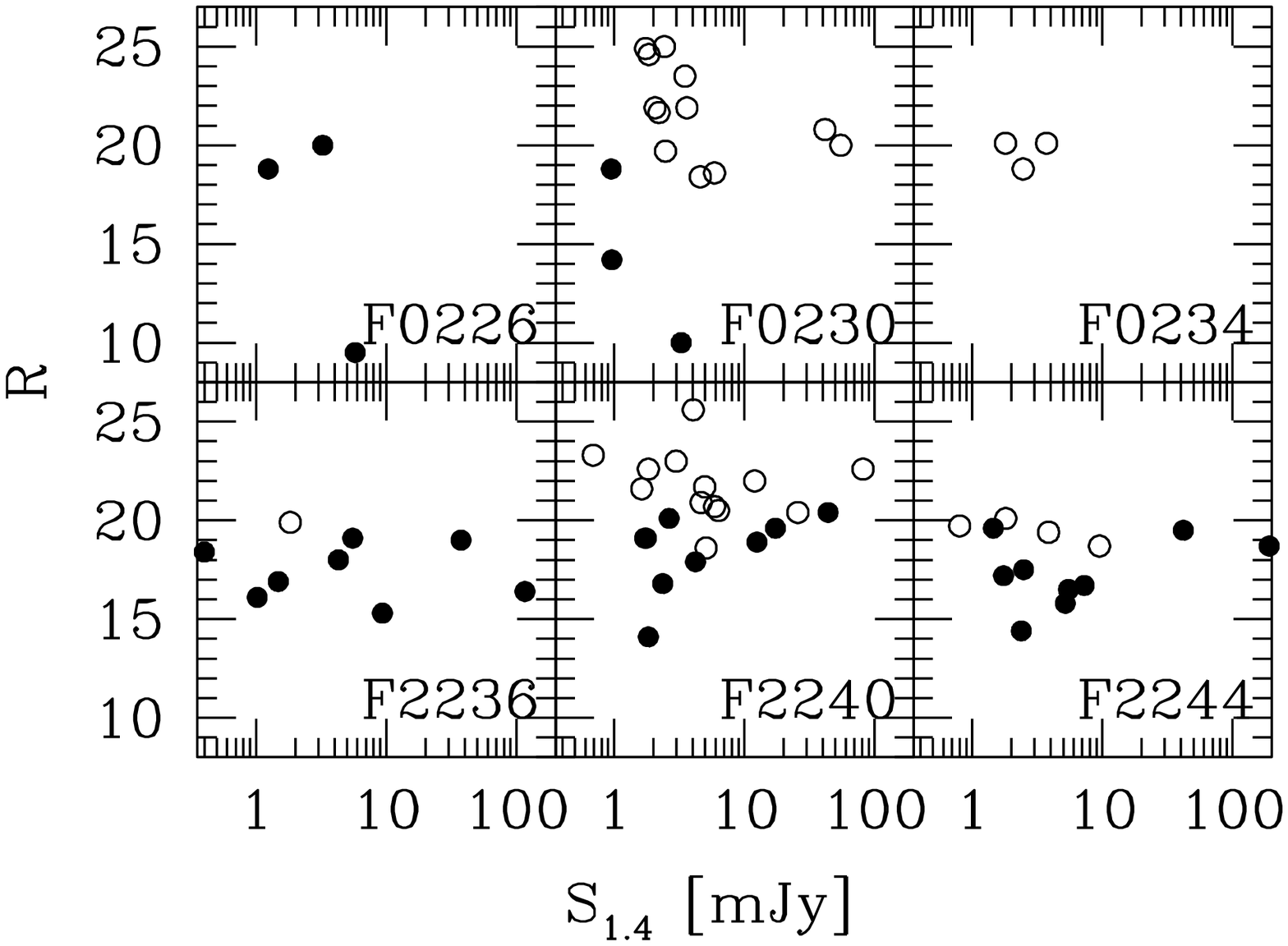}
\caption{Apparent magnitude R vs. radio flux S$_{1.4 GHz}$ for the
  objects in the fields as shown in the figure. Filled dots indicate
  sources with a redshift determination, while circles are
  for objects with no $z$.
\label{fig:mags_flux} }
\end{figure*}

\subsection{The APM Survey}

Some of the fields (F0226, F0230, F0234, F2236, F2240, F2244) targeted
in our WHT observations lie in regions of the sky sampled by the APM
scan of UKST plates ({\tt http://www.ast.cam.ac.uk/$^\sim$apmcat/}).
The APM data provided R and B magnitudes of the objects in the sample
brighter than B $\sim 22$, R $\sim21$, the limiting magnitudes of
the APM survey.

To obtain these magnitude measurements we searched the UKST catalogue
for optical counterparts lying within 2.7 arcseconds from the
corresponding radio position. Note
that the fibre position always coincides with the radio coordinates
except in the case of double sources (see Section 3.3). 
The tolerance in  positional matching allows for measurement errors
and the  fact that the centre of radio emission might be displaced
from the centre of the optical emission, especially if the radio object is
extended.  If no optical object within this offset was found in the
UKST catalogue we assumed the object to be fainter than the limiting
magnitude.

A few objects in the F0230 and F2240 fields show a displacement
greater than the value of 2.7${''}$ (but within 3$''$ - Table~2). 
These are the sources observed at the MDM Observatory (see 
Section~\ref{sect:further}).

Figure \ref{fig:mags_flux} plots the apparent magnitude R vs. the
radio flux S$_{1.4\,{\rm GHz}}$ for all those sources with an optical
counterpart in the APM catalogue. Filled dots represent objects with
a redshift determination, while circles are for those objects with
no $z$. The plot indicates that the
percentage of successful redshift determinations is $100\%$ to R $\sim
18.6$ and then decreases with increasing apparent magnitude at a rate
which varies from field to field. 
Figure \ref{fig:mags_flux} also shows that almost all the objects in
our spectroscopic sample
with a redshift determination are brighter than the APM limiting
magnitude; only 3 sources with a measured redshift are not detected in
the APM data. 

\begin{figure*} 
\vspace{16cm}  
\includegraphics{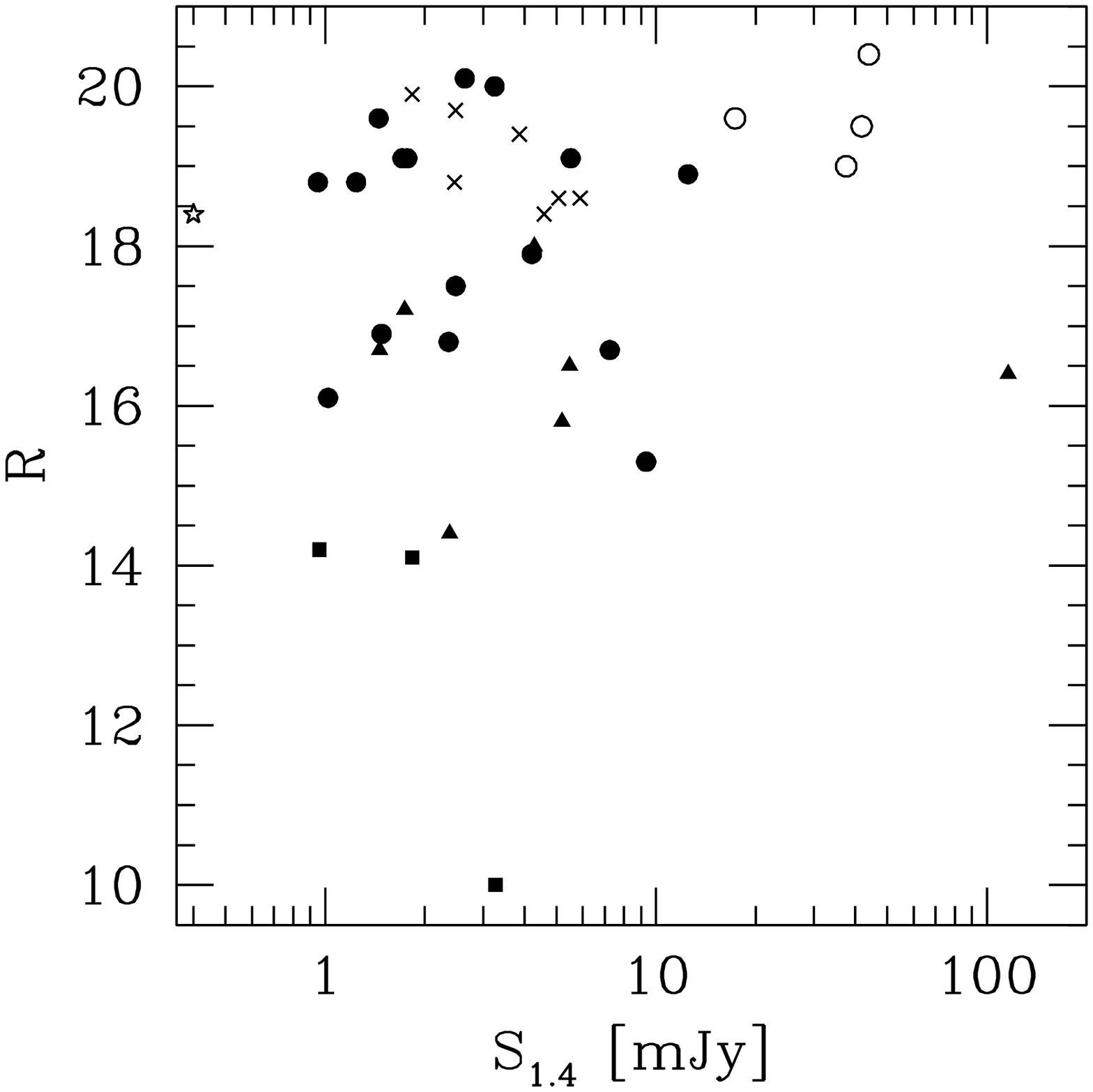}
\includegraphics{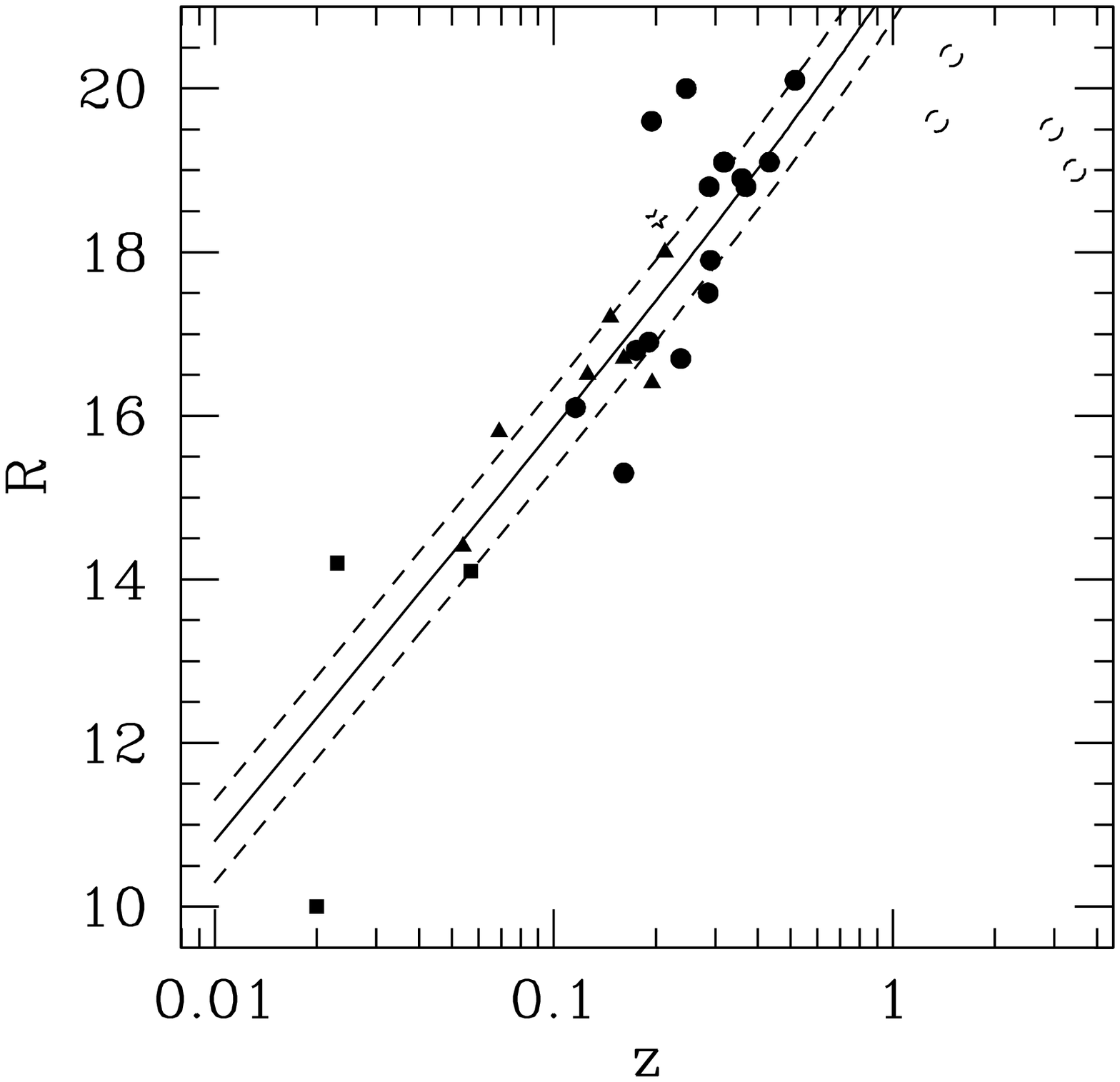}
\includegraphics{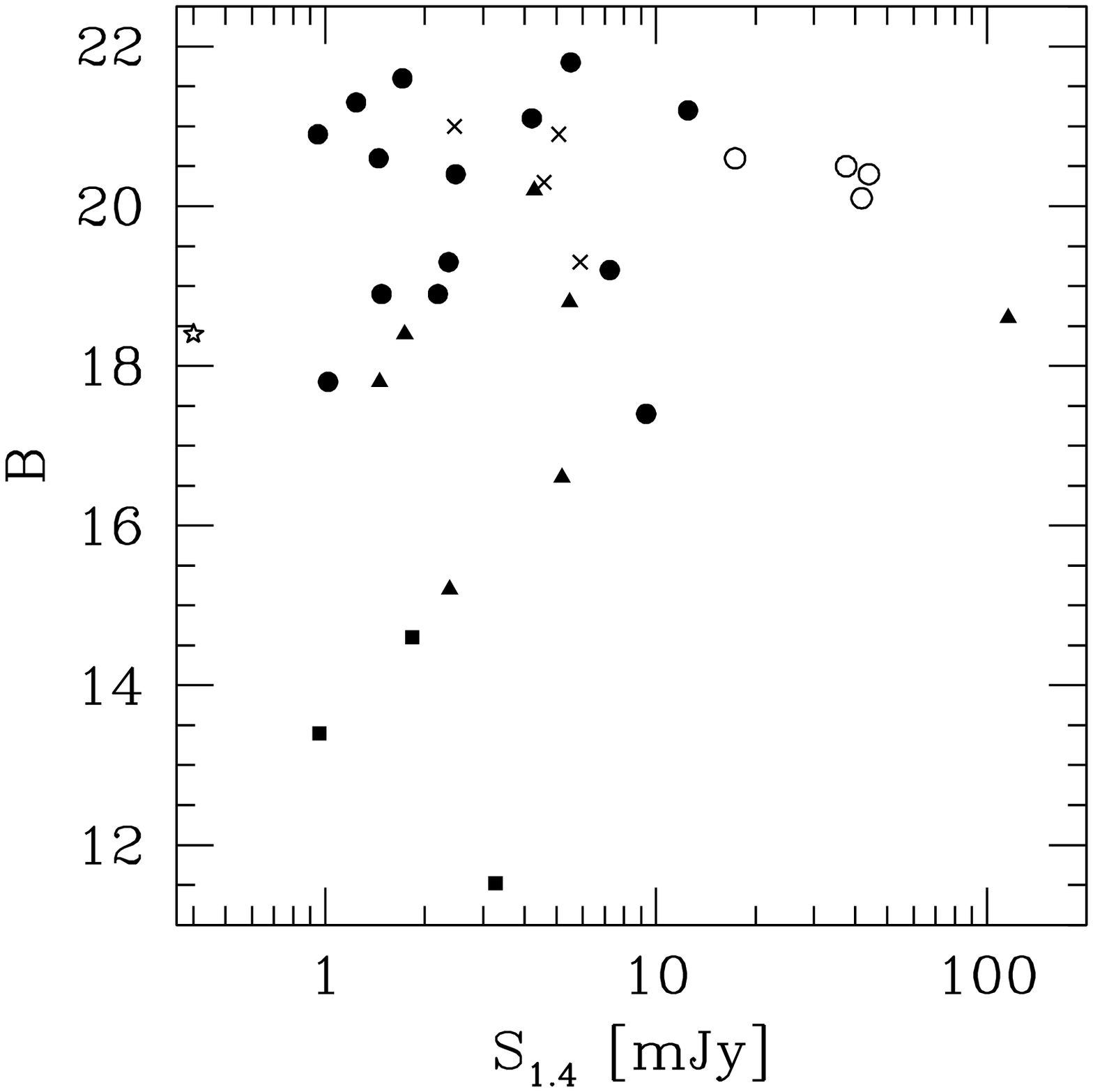}
\includegraphics{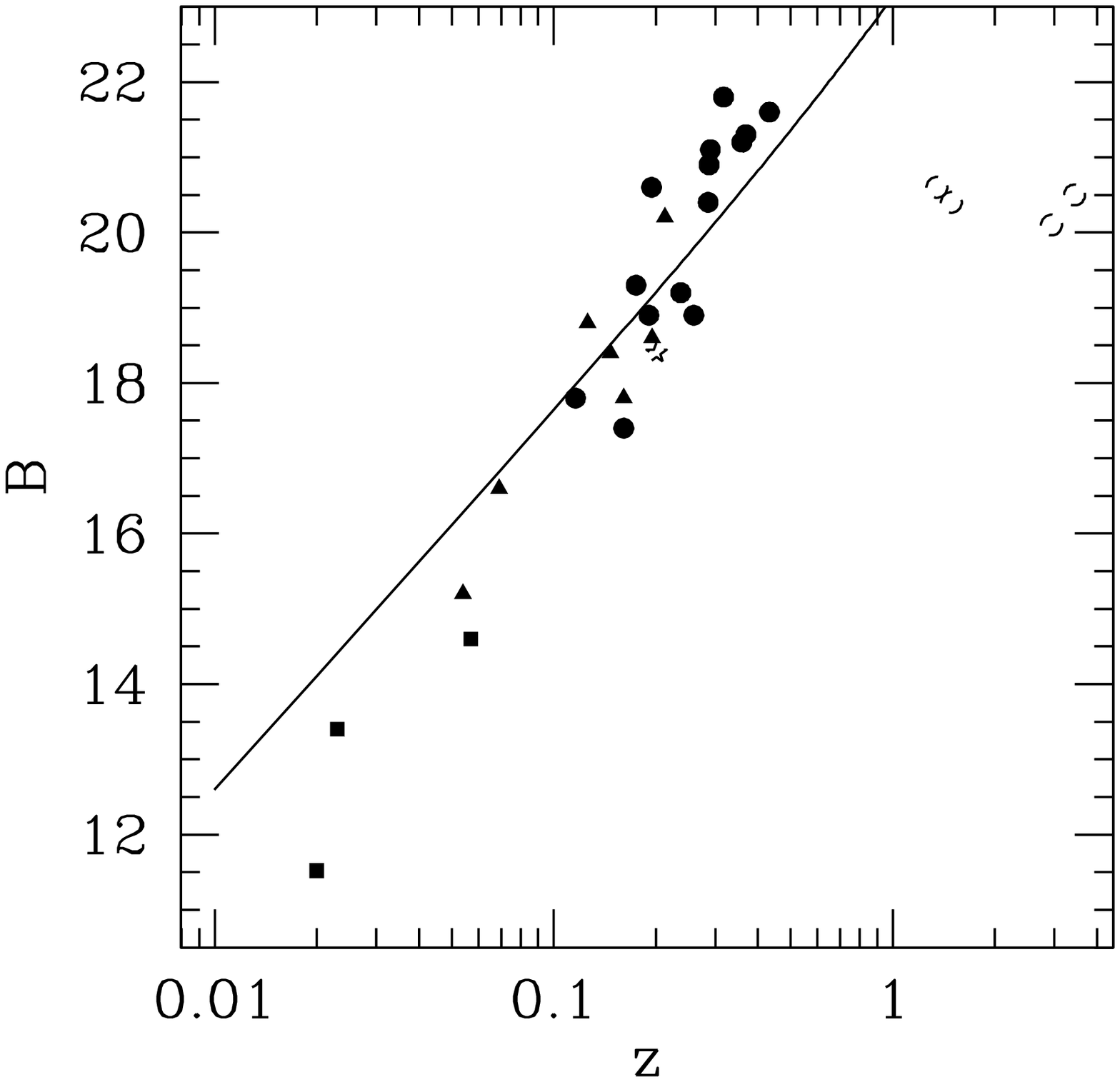}
\caption{R magnitude versus radio flux at 1.4 GHz (top-left) and
  redshift $z$ (top-right); B magnitude versus radio flux at 1.4 GHz
  (bottom-left) and redshift $z$ (bottom-right). Filled dots are for
  early-type galaxies, filled triangles for late-type galaxies, filled
  squares for starburst galaxies, open dots for
  Seyfert1 galaxies / quasars, and stars indicate Seyfert2 galaxies. Crosses correspond to objects with no redshift measurement.
The solid lines in the R-$z$ and B-$z$ plots represent the best fit to the
  data (see text), while the dashed lines in the top-right
  panel show 2$\sigma$ confidence limits. 
\label{fig:plots}} 
\end{figure*}

\subsection{Further observations}
\label{sect:further}
CCD observations at the MDM Observatory have yielded R~magnitudes
($21<{\rm R}<26$) for objects in the F0230 and F2240 fields as shown in
Figure \ref{fig:mags_flux} and Table~\ref{Table_OIDs} (D. Helfand,
private communication).

\subsection{Field-to-field scatter}
The success rate of redshift determination varied strongly from field
to field as shown in Table~\ref{Table_fields}; $N_{obs}$ and $N_{z}$ are
respectively the number of  observed sources and the number of sources
with redshift determinations. The striking differences in redshift
success between fields are illustrated in Figure \ref{fig:mags_flux};
the fields at 22$^h$ have far more redshift determinations than the
fields at 02$^h$, which average only a $\sim 5$ per cent success rate. 
This region also has an anomalously low number of optical
identifications in the FIRST survey (D. Helfand private
communication), even though the surface density of radio sources is
close to the average. 

Comparing the two fields for which we have photometric information
down to R $\sim 26$, in the case of F2240, E(B-V)=0.067, while for
F0230, E(B-V)=0.027 (D. Helfand, private communication). Thus it is
not Galactic extinction that produces the very different
identification rates. We are looking at either intergalactic
extinction or evidence of large-scale structure.

If we assume that radio galaxies cluster as optical galaxies, we can
make a crude estimate of the expected variance in the number of
measured redshifts in each WHT field.  As discussed later, the typical
redshift of galaxies in our sample is $z\sim 0.2$, and so each
1$^\circ$ field samples a volume of roughly $1.7\times 10^{4} h^{-3}$
Mpc$^3$ out to $z=0.2$. Ignoring the difference in shape, this volume
is equivalent to a 26$h^{-1}$Mpc cube, for which we would expect a
fractional variance in the number of optical galaxies to be between
0.3 and 0.4 (Loveday et al., 1992). The observed field-to-field
variance is 0.6, which may reflect the higher clustering amplitude of
radio galaxies.

Whatever the true explanation, the observed variation is important in
itself because of implications for the number of radio-sources with
faint optical continuum but bright emission lines. We return to this
issue in Section~8 when estimating $N(z)$.

\section{OPTICAL AND RADIO PROPERTIES OF THE SAMPLE}
Table~\ref{Table_OIDs} shows the properties of objects in our sample
for which we have data (either spectroscopic or photometric) on the optical
counterpart. The columns are as follows:\\ 
(1) Source number.\\ 
(2) Right ascension and declination of the targeted object. These
coordinates are from those of the FIRST survey except in the case
of 2240\_004 for which the fibre was placed at the centroid of a
double source.\\ 
(3) Offset of the optical counterpart in the APM
catalogue before correction for sytematic offsets.\\
(4-5) Apparent magnitudes R and B of the optical
counterpart.\\ 
(6) Radio flux density at 1.4 GHz.\\ 
(7) Redshift.\\ 
(8) Spectral classification.\\ 
(9) Emission lines detected.

Plots for the subsample with redshift determinations  
are presented in  Figure \ref{fig:plots} for
(1) R magnitude vs. radio flux, (2) R vs. redshift, (3) B magnitude vs.
radio flux, and (4) B vs. redshift.  
The solid line in the R-$z$ plot of Figure \ref{fig:plots} 
represents the relation
\begin{eqnarray}
R-M_R=-5+5{\rm log}_{10}d_L (pc)\simeq
25-5{\rm log}_{10}{\rm H}_0+\nonumber \\+5{\rm log}cz+1.086 (1-q_0)z
\end{eqnarray}
($d_L$ the luminosity distance), obtained by a minimum $\chi^2$ fit
to the data with the absolute magnitude $M_R$ as free parameter. 
This analysis, for $H_0=$~50~Km~sec$^{-1}$Mpc$^{-1}$ and $q_0=0.5$,
gives a value of $M_R\simeq - 23.1$ with little scatter ($\Delta
M_R\sim 0.5$ at the $2\sigma$ level - dashed lines in Figure
\ref{fig:plots}).
There are clearly two populations in the diagram, and to find the best
fit for early types we excluded Seyfert-type galaxies and starbursts;
the result from using all galaxies was virtually identical. The 
result is in very good agreement with previous estimates (see
e.g. Rixon, Wall \& Benn, 1991; Gruppioni, Mignoli \& Zamorani, 1998;
Georgakakis et al., 1999), showing that passive radio galaxies are
reliable standard candles.  The solid line in the B-$z$ plane of
Figure \ref{fig:plots} shows a $\chi^2$~fit yielding $M_B\simeq
-21.3$, although the fit is not as good as that in the R-$z$
plane. The fit is improved if only early-type galaxies are included.

\begin{figure}
\vspace{7cm}  
\includegraphics{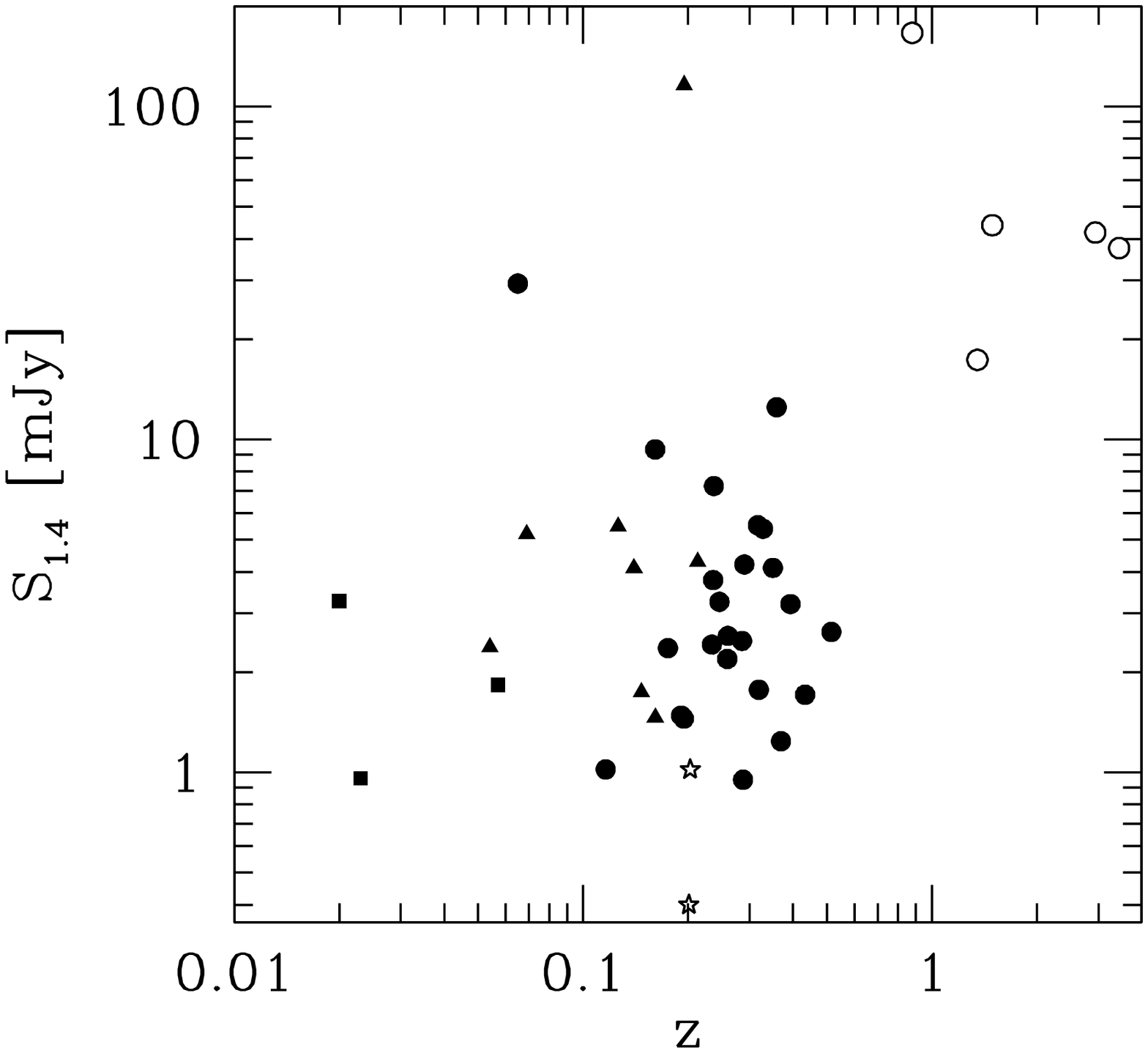}
\caption{Radio flux at 1.4 GHz versus redshift. Filled dots are for
  early-type galaxies, filled triangles for late-type galaxies, filled
  squares for starburst galaxies, open dots for
  Seyfert1 galaxies / quasars, and stars indicate Seyfert2 galaxies.
\label{fig:flux_red}} 
\end{figure}

Figure \ref{fig:flux_red} shows radio flux S$_{1.4 \rm{GHz}}$
vs. redshift for the objects with a redshift determination.  Though
the number of objects in each class is small, we can see that
different classes of objects occupy different regions of the planes in
Figures \ref{fig:plots} and \ref{fig:flux_red}:
\begin{itemize}
\item Early-type galaxies are found at intermediate redshifts ($0.1
  \la z\la 0.6$) and make up the entire spectroscopic sub-sample with
  $0.2\la z \la 0.5$. Their radio fluxes lie in the range $1\la {\rm
  S}_{1.4\,{\rm GHz}} \la 10$ mJy and optically they appear as
  relatively faint objects (R $\ga 17$, B $\ga 18.5$). 
  Their absolute magnitude $M_R\sim -23$ is consistent with
  FRI-type radio galaxies (Ledlow \& Owen, 1996).
  

\item Late-type galaxies are found from low to intermediate redshifts
  ($0.05\la z \la 0.2$) perhaps constituting an intermediate step
  between early-type and starburst galaxies. Their radio fluxes range
  from S$_{1.4 \rm{GHz}}\simeq 1.5$ mJy to S$_{1.4\,{\rm GHz}}\simeq 6$
  mJy, with a single exception of one radio-bright object (2236\_013,
  S$_{1.4 \rm{GHz}}=116$ mJy).  The B and R magnitudes show intermediate
  values, $15\la {\rm R}\la 17 $ and $15\la {\rm B}\la 19$.

\item Starburst galaxies in the sample are very nearby ($z\la 0.05$),
  optically-bright (R $\la 14$, B $\la 14.5$), and faint in radio
  emission (S$_{1.4\,{\rm GHz}}\la 3$ mJy). Two of them (0230\_071 and
  0230\_061) are separated by  $\la 1 h^{-1}$Mpc and show very
  similar properties, suggesting that interaction may be responsible for the
  enhanced star-formation rate. 

\item Seyfert2 galaxies are found at intermediate redshifts ($z\sim
  0.2$) and their radio flux densities are extremely low (S$_{1.4\,{\rm
  GHz}}\la 1$ mJy). Photometric data are available for only one. These
  objects lie close to the line delimiting Seyferts from HII galaxies in
  the log(OIII/$H\beta$)/log(OII/$H\beta$) plane (Rola, Terlevich \&
  Terlevich, 1997).

\item Seyfert1 galaxies / quasars exhibiting broad emission lines are
  only found at $z\ga 0.9$. These objects are very bright in the radio
  but quite faint in both the R and B bands.
  2240\_004 is a double source showing characteristic radio lobes
  symmetric with respect to the optical centre (which does not show a
  radio core).

\item Unclassified objects. These are probably early-type galaxies
  since they occupy the region in the R-S and B-S planes which is
  occupied by early-type objects. Also their colours are consistent with
  them being red/old objects; almost 50~per~cent of objects this
  subsample have R magnitudes (with $18.6\la {\rm R}\la 20$) in the APM
  catalogue but no B magnitudes, because they are below the APM survey
  limit.  For these objects we used the R-$z$ relation of Figure
  \ref{fig:plots} to estimate redshifts. The corresponding values are
  given in Table~\ref{Table_OIDs}, distinguished by asterisks. (Note
  that we do not use these redshift estimates in the following
  analysis.)

\end{itemize}

\begin{table*}
\begin{center}
\caption{Observed FIRST sources. Where an APM optical identification
was found the offset from the radio position, and R and B magnitudes
are listed. The measured redshifts and spectral type are listed for the
targets detected in the spectral data. \label{Table_OIDs}} 
\begin{tabular}{llllllllll}
Object& $\alpha$ (J2000)& $\delta$ (J2000)  & Offset [$\prime$$\prime$]&
R & B& S$_{1.4}$ [mJy]& z& Class& Emission Lines\\
\hline
 0226\_041 & 2 29 53.551  & +00  2 18.53   & $0.68$ &$18.8$  &$21.3$
&$1.24$ &$0.369$ &Early& none\\
 0226\_068  & 2 30 21.753  & +00  5 11.21   &$2.66 $ &$9.5 $  &$11.4$  &$5.77$ &$0.0$   &Star&none\\
 0226\_063  & 2 30 14.969  & +00  22 39.30  & $0.59$ &$20.0$  &$>22$
&$3.25$ &0.246     &Early& OII\\
 0230\_048  & 2 33  5.021  & +00 32 39.60   & ? &$>26  $ &$>22  $ &$3.2$  &$0.393$ &Early&OII\\
 0230\_071  & 2 32 47.492  & +00 40 40.85   &$0.38$ &$14.2$  &$13.4$
&$0.96$ &$0.023$ &SB&H${\gamma}$,H${\beta}$,OIII,H${\alpha}$\\
 0230\_063  & 2 31 28.228  & -00  7 34.35   &$0.53$ &$18.8 $ &$20.9$  &$0.95$ &$0.2873$&Early&OII\\
 0230\_061  & 2 31 37.698  & -00  8 23.57   &$0.18$ &$10.0 $ &$11.5$
&$3.27$ &$0.02$  &SB&H${\gamma}$,H${\beta}$,OIII,H${\alpha}$ \\
 0230\_018  & 2 33  2.830  & +00 24  5.90   & ?  &$>26  $  &$>22 $  &$3.8$   & ?     &?\\
 0230\_006  & 2 32 47.092  & +00 19 50.46   &$2.8$  &$25$    &$>22 $  &$2.43$  &?    &?\\
 0230\_029  & 2 32 28.230  & +00 27 44.27   &? &$>26  $ &$>22 $  &$3.08$  &?    &?\\
 0230\_066  & 2 31 57.592  & +00 38 54.68   &$0.71$ &$19.7$  &$>22$
&$2.48$  &$\sim 0.53^{\ast}$   &Unclass& none\\
 0230\_037  & 2 32  6.979  & +00 28 45.46   &$1.3 $ &$23.5 $ &$>22  $ &$3.5  $ &?    &?\\
 0230\_023  & 2 32  7.959  & +00 25 26.69   &?  &$>26$   &$>22$   &$1.48$  &?    &?\\
0230\_009 &2 32  9.710 &+00 22 36.90  &? &$>26$   &$>22$   &28.56 &?    &?\\
0230\_054 &2 31  5.596 &+00  8 44.00  &0.91 &20    &20.7  &55.01 &?   &?\\
0230\_002 &2 32 29.895 &+00 12  1.97  &?   &$>26$   &$>22$   &1.47  &?    &?\\
0230\_031 &2 32  9.524 &-00  1 28.66  &?   &$>26$   &$>22$   &1.24  &?    &?\\
0230\_041 &2 32 16.497 &-00  4 45.56  &?   &$>26$   &$>22$   &3.10  &?    &?\\
0230\_008 &2 32 28.662 &+00  2 33.72  &$1.4$  &24.9  &$>22$   &1.74  &?    &?\\
0230\_025 &2 32 32.444 &-00  0 44.07  &2.4  &21.7 &$>22$   &2.21  &?    &?\\
0230\_067 &2 33 13.848 &-00 12 14.83  &1.11 &18.6  &19.5   &5.9
&$\sim 0.33^{\ast}$&Unclass&none\\
0230\_010 &2 32 52.548 &+00  2 34.79  &1.4  &20.8  &$>22$   &41.56 &?    &?\\
0230\_014 &2 33  0.332 &+00  2 35.44  &?   &$>26$   &$>22$   &0.69  &?    &?\\
0230\_013 &2 33  8.894 &+00  4 42.19  &$1.8$  &24.6  &$>22$   &1.85  &?    &?\\
0230\_028 &2 33 31.030 &+00 10 23.61  &0.94  &18.4  &20.3  &4.59
&$\sim 0.31^{\ast}$    &Unclass&none?\\
0230\_026 &2 33 30.170 &+00 14 48.03  &1.4  &21.9  &$>22$   &2.06  &?    &?\\
0230\_011 &2 33 15.494 &+00 18 34.54  &0.0  &21.9  &$>22$   &3.62  &?    &?\\
0234\_026 &2 35 11.618 &+00  6  8.29  &?  &$>21$   &$>22$   &5.39 &0.328&Early&none\\
0234\_051 &2 37  3.870 &+00 39 23.88  &0.9 &20.1  &20.5  &1.8  &?   &?\\
0234\_005 &2 36  9.341 &+00 18 57.04  &0.28 &18.8  &21.0  &2.46
&$\sim0.36^{\ast}$   &Early&none?\\
0234\_041 &2 34 57.578 &+00 21 30.49  &0.41 &20.1  &21.0  &3.72 &?   &?\\
2236\_005 & 22 38 45.422 &+00  8 39.11& 0.23 & 19.1 &  21.8  & 5.52 & 0.317 & Early&OII \\
2236\_012 & 22 39 32.234 &+00 12 42.49&  1.77  & 16.7 &  17.8  & 1.46 &
0.161 & Late& H${\beta}$,OIII\\
2236\_001 & 22 38 43.562 &+00 16 48.15&  0.29 & 19.0 &  20.5  &37.54
& 3.44  & Sy1& Ly${\alpha}$,CIV\\
2236\_023 & 22 39  8.156 &+00 32 32.09&  0.79 & 18.0 &  20.2  & 4.29
& 0.213 & Late& OIII\\
2236\_008 & 22 38  2.438 &+00 23 34.55&  0.1 & 18.4 &  20.2  & 0.40
& 0.201 & Sy2&OII,OIII\\
2236\_013 & 22 37 37.117 &+00 20 39.02&  0.1 & 16.4 &  18.6 &115.99
& 0.195 & Late& OII,OIII\\
2236\_009 & 22 37 49.289 &+00 19  0.85&  0.22 & 16.1 &  17.8  & 1.02 & 0.116 & Early&none\\
2236\_034 & 22 36 51.883 &+00 12 23.11&  0.72 & 16.9 &  18.9  & 1.48 & 0.191 & Early&none\\
2236\_016 & 22 37 28.898 &+00 10  1.02&  ?  & $>21$  & $ >22$   & 66.11& 0.0   & Star&none\\
2236\_046 & 22 37  6.594 &-00  2 31.66&  1.2 & 15.3 &  17.4  & 9.33 & 0.161 & Early&none\\
2236\_028 & 22 39 18.391 &-00  4 34.18&  0.23 & 19.9 &  $>22 $  &
1.83 &$\sim 0.50^{\ast}$&Unclass&none?\\
2240\_002 & 22 42 22.117 &+00 15  5.80&  0.3 & 19.6 &  20.6  &17.33 &
1.35  & Sy1&CIII,MgII \\
2240\_008 & 22 42 12.805 &+00  8 12.71&  0.69  & 20.1 &  $>22$   & 2.64 & 0.515 & Early&none\\
2240\_023 & 22 42 12.055 &+00  2 11.83&  0.81 & 19.1 &  21.6  & 1.71 & 0.433 & Early&none\\
2240\_004 & 22 42 43.141 &+00 11 17.62&  ? & $>26$ & $>22$  & 4.11  &0.35  & Early&none\\
2240\_038 & 22 44  4.414 &+00 12 16.71&  0.43 & 18.9 &  21.2  &12.49 & 0.359 & Early&none\\
2240\_016 & 22 43 22.695 &+00 15 21.01&  0.5  & 17.9 &  21.1  & 4.21 & 0.29  & Early&none\\
2240\_033 & 22 43 38.438 &+00 25 17.64&  1.82 & 16.8 &  19.3  & 2.36 & 0.175  &Early&OII,OIII\\
2240\_021 & 22 43 15.477 &+00 24 28.93&  1.43 & 14.1 &  14.6&   1.83 & 0.057 & SB&H${\gamma}$,H${\beta}$,OIII,H${\alpha}$\\
2240\_010 & 22 42 27.078 &+00 25 57.40&  0.55 & 19.1 & $>22$ &  1.77 & 0.319 & Early&none\\
2240\_014 & 22 41 46.367 &+00 16  8.19&  0.54 & 20.4 &  20.4 & 43.96 &
1.49  & Sy1&CIII,MgII\\
2240\_040 & 22 41 45.398 &-00  4  1.12&  1.9  & 25.6 &  $>22 $ &  4.05 &  ?    & ?\\
2240\_011 & 22 42 15.273 &+00  6 19.84&  0.0  & 21.7 &  $>22$   & 4.96 &  ?    & ?\\
2240\_050 & 22 41 46.234 &-00  8 44.18&  ?  & $>26$  & $ >22$   & 2.96 &  ?    & ?\\
2240\_042 & 22 42 10.977 &-00  6 57.91&  1.4  & $>26 $ & $ >22 $  & 2.99 &  ?    & ?\\
2240\_006 & 22 42 33.227 &+00  8 58.03&  ?   & $>26$  & $ >22$   & 1.30 &  ?    & ?\\
2240\_005 & 22 42 51.148 &+00 11 43.35&  1.78  & 18.6 &  20.9  & 5.08 & $\sim0.33^{\ast}$&Unclass&none\\
2240\_025 & 22 43 34.125 &+00  8 46.66&  ?   & $>26$  & $ >22$  &  3.40 &  ?    & ?\\
2240\_024 & 22 43 38.047 &+00 17 49.9&  3.0 &  22.6  & $>22 $ & 81.53  & ?     &?\\
2240\_015 & 22 43 19.602 &+00 19 17.92&  2.4 &  20.9 &  $>22 $ &  4.66 &  ?    & ?\\
2240\_019 & 22 43 17.797 &+00 22 53.89&  ?& $ >26$  & $ >22$  &  3.07 &  ?    & ?\\
2240\_034 & 22 43 35.289 &+00 27 32.72&  1.2 &  23.3 &  $>22$  &  0.69 &  ?    & ?\\
2240\_022 & 22 43 10.891 &+00 26 54.05&  ?  & $ >26$  & $ >22$  &
0.99 &  ?    & ?\\
\end{tabular}
\end{center}
\end{table*}
\begin{table*}
\begin{center}
\contcaption{}
\begin{tabular}{llllllllll}
  Object& $\alpha$ (J2000) & $\delta$ (J2000)  & Offset [$^{\prime\prime}$]& R & B&
S$_{1.4}$ [mJy]& z& Class&Emission Lines\\
\hline
2240\_036 & 22 43 25.695 &+00 33 20.03&  0.5 &  22.6 &  $>22$  &  1.83 &  ?    & ?\\
2240\_049 & 22 43 33.211 &+00 37 51.37&  0.18 &  20.4 & $ >22$  & 25.74 &  ?    & ?\\
2240\_001 & 22 42 31.938 &+00 17 41.52&  2.2 &  21.6 &  $>22 $ &  1.63 &  ?    & ?\\
2240\_009 & 22 42 23.633 &+00 24 41.88&  ? &  $>26$  &  $>22$  &  6.95 &  ?    & ?\\
2240\_041 & 22 41 44.898 &+00 35 26.75&  0.45&  20.5 & $ >22$  &  6.33 &  ?    & ?\\
2240\_037 & 22 41 25.992 &+00 31  4.84&  ? &  $>26$  &  $>22$  &  0.55 &  ?    & ?\\
2240\_012 & 22 42  2.547 &+00 22 54.23&  0.3 &  20.7 &  $>22$&    5.91 &  ?    & ?\\
2240\_030 & 22 41 30.938 &+00 24 41.72&  2.18 &  20.5 & 21.6&  12.00 &  ?    & ?\\
2244\_039 & 22 47  5.680 &+00 35 31.59&  0.31&  19.6 &  20.6&   1.45 & 0.1945& Early&none\\
2244\_007 & 22 46 41.789 &+00 28 36.51&  1.26&  17.5&   20.4&   2.48 & 0.2855& Early&none\\
2244\_013 & 22 45 52.422 &+00 27 52.05&  0.54&  16.7 &  19.2&   7.24 & 0.237 & Early&none\\
2244\_068 & 22 45 18.352 &+00 34 45.15&  0.37&  $>21$  &  21.6&   6.73 &  ?    & ?\\
2244\_006 & 22 45 54.953 &+00 23 10.08&  0.11&  15.8 &  16.6&   5.19
& 0.069 & Late&H${\gamma}$,H${\beta}$,OIII,OI,H${\alpha}$\\
2244\_029 & 22 45 32.836 &+00 26 40.19&  0.24&  16.5 &  18.8&   5.48 & 0.126 & Late&H${\beta}$,OIII\\
2244\_077 & 22 44 59.453 &+00  0 33.55&  0.38&  19.5&   20.1&  41.82
& 2.94  & Sy1&Ly${\alpha}$,CIV\\
2244\_044 & 22 45 59.062 &-00  4 14.14&  0.49&  17.2 &  18.4&   1.74 & 0.147 & Late&H${\beta}$,OIII\\
2244\_061 & 22 47 55.789 &+00  0 12.49&  1.99&  17.2 &  18.5&    0   & 0.0    &Star&none\\
2244\_086 & 22 48 24.617 &+00  9 21.06&  0.34&  14.4 &  15.2&   2.38 & 0.054 & Late&H${\gamma}$,H${\beta}$,OIII,H${\alpha}$\\
2244\_008 & 22 45 40.062 &+00 10 53.50&  1.00&  19.4 &  $>22$ &
3.86 & $\sim 0.45^{\ast}$  &Unclass&OII?\\
2244\_027 & 22 45 41.258 &+00  2 54.64&  0.51&  19.7&  $>22$ &   0.80 &  ?   &  ?\\
2244\_038 & 22 47 30.195 &+00  0  6.68&  0.2&  18.7 &  19.9& 190.56 & 0.0  &  Star&none\\
2244\_049 & 22 47 54.352 &+00  4 38.91&  1.22&  18.7 &  $>22$ &   9.49 &  ?   &  ?\\
2244\_023 & 22 46 11.922 &+00 32 32.35&  0.85&  20.1 &  20.6&   1.81
&  ?   &  ?\\
2244\_052& 22 46 53.570 &-00 8 7.86& 2.01& $>21$&18.9& 2.19&0.259&Early&OII\\
ELAIS\_059& 16  9 23.250 &+54 43 22.42&    ? &   ?   &   ? &    3.78 & 0.236&  Early&none\\
ELAIS\_032& 16  8 58.008 &+54 18 17.83&    ? &   ?   &   ? &    2.57 & 0.260&  Early&none\\
ELAIS\_060& 16  8 28.340 &+54 10 30.81&    ? &   ?   &   ? &    2.42 & 0.234&  Early&OII\\
ELAIS2\_043& 16  1 28.094& +53 56 50.27&   ? &   ?   &   ? &   29.39 & 0.065&  Early&none\\
ELAIS2\_011& 16  5 15.977& +53 54 10.53&   ? &   ?   &   ? &    1.02 & 0.203&  Sy2&OII,NeIII,H${\beta}$,OIII\\
ELAIS2\_017& 16  5 52.664& +54  6 51.03&   ? &   ?   &   ? &    4.10 & 0.14 &  Late&H${\delta}$,H${\gamma}$,H${\beta}$,OIII\\
ELAIS2\_024& 16  6 23.555& +54  5 55.66&   ? &   ?   &   ? &  166.23 & 0.878&  Sy1&MagII\\
\end{tabular}
\end{center}
\end{table*}

\section{THE REDSHIFT DISTRIBUTION $N(z)$}

To construct the redshift distribution of radio sources $N(z)$ at
the 1-mJy level we use the WYFFOS data described here (the WHT sample) 
and measurements from the Phoenix survey (Hopkins et al., 1998). 

\subsection{The WHT Sample}

\begin{figure} 
\vspace{8cm}  
\includegraphics{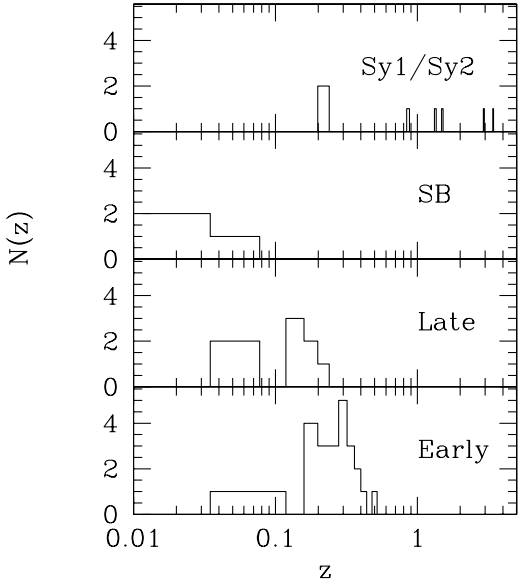}
\caption{Redshift distribution of objects in our sample according to
spectral type.
\label{fig:N_z1}} 
\end{figure}

The number of objects in the WHT sample belonging to each
spectroscopic class (Section 6) are summarized in
Table~\ref{Table_types}.
\begin{table}
\caption{The number and fraction of each spectroscopic class. 
\label{Table_types} }
\begin{tabular}{lll}
  Type& Number of Objects& \% of the $z$-sample  \\
\hline
Early &24 &45 \\
Late &8&15 \\
Starburst& 3&6 \\
Seyfert1 &5&9 \\
Seyfert2 &2&4 \\
Unclassified & 7&13\\
Stars & 4&7\\
& &\\
\end{tabular}
\end{table}
The redshift distribution of these objects split by 
their spectral type is presented in Figure \ref{fig:N_z1}. Here
AGN-powered sources (i.e. Seyfert1 /quasars and Seyfert2) are plotted
together, the low-$z$ objects corresponding to Seyfert2's.
Our results are in good agreement with the predictions of Jackson \&
Wall (1998) for the relative contribution of different classes of
objects at mJy levels (see Figure 17 of Jackson \& Wall, 1998).  At
S$_{1.4\,{\rm GHz}}\ge 1$~mJy their models predict (C.A. Jackson, private
communication) 6.4~per~cent of the whole population to be broad- and
narrow-emission line objects (high excitation FRII's, comparable with
our class of Seyfert1+Seyfert2 galaxies), 48.1~per~cent to be
low-excitation FRI's and FRII's (early-type galaxy spectra), 28.7~per~cent
to be starforming galaxies (i.e. starbursting+late type) and a final
16.8~per~cent to be BL Lacs (no features in their spectra - some of which
may be amongst our unclassified objects). 

\begin{figure} 
\vspace{8cm}  
\includegraphics{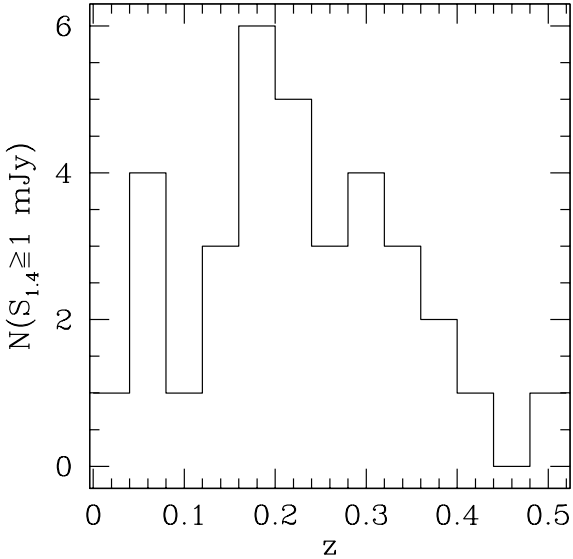}
\caption{Redshift distribution of objects in our sample with
  S$_{1.4\,{\rm GHz}}\ge1$~mJy.
\label{fig:N_z}} 
\end{figure}

Figure \ref{fig:N_z} shows the redshift distribution of all
extragalactic objects of our sample with S$_{1.4\,{\rm GHz}}\ge 1$~mJy
and $z\la 0.6$.  These flux-density and redshift limits cut 8 objects
from the spectroscopic sample (all 5 Seyfert1 galaxies, one
early-type, one Seyfert2 and one starburst galaxy), so that Figure
\ref{fig:N_z} contains a total of N$_{\rm TOT} = 34$ sources.


\subsection {The Phoenix Sample}
\begin{figure} 
\vspace{8cm}  
\includegraphics{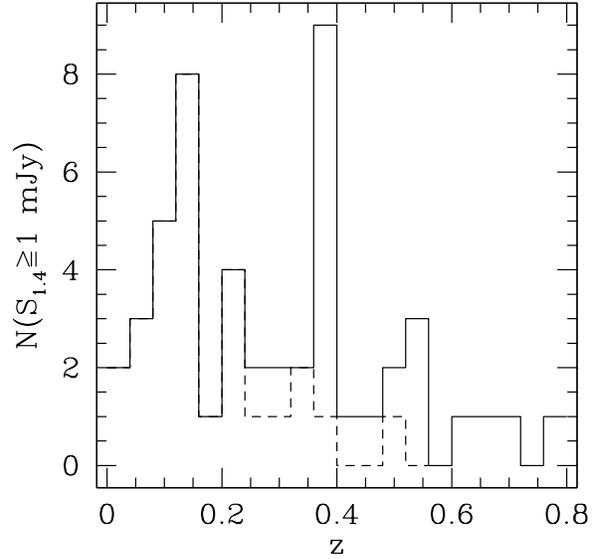}
\caption{Redshift distribution for sources 
 in the Phoenix survey with S$_{1.4\,{\rm GHz}}\ge 1$~mJy (solid line). 
The dotted line shows the 
  distribution obtained by imposing the further constraint $R\le 18.6$.
\label{fig:N_zp}} 
\end{figure}

The Phoenix survey at 1.46 GHz (Hopkins et al., 1998) includes
observations from two different fields. The Phoenix Deep Field (PDF)
covers an area $2^\circ$ in diameter centred at RA(2000)=$01^h\;
14^m\; 12^s.16$, Dec(2000)=$-45^\circ\;
44^{\prime}\;8^{\prime\prime}.0$, while the Phoenix Deep Field
Sub-region (PDFS) covers an area of $36^{\prime}$ in diameter centred at
R(2000)=$01^h\;11^m\;13^s.0$, Dec(2000)=$-45^\circ\; 45^{\prime}\;
0^{\prime\prime}.0$. The completeness of the radio catalogue is
estimated to be $\sim$~80~per~cent down to 0.4 mJy for the PDF
catalogue and $\sim$~90~per~cent down to 0.15 mJy for the PDFS
catalogue. Optical and near-infrared data were obtained for many radio
sources in the sample. A catalogue of 504 objects with redshift
estimates and photometric information down to a limiting magnitude of
R=22.5 has been produced. This represents $\sim$~47~per~cent of the
whole radio sample. From this catalogue, the properties of the sub-mJy
population have been analysed (Georgakakis et al., 1999).

To compare directly with our data, we have selected only objects with
radio fluxes S$_{1.4\,{\rm GHz}}\ge 1$ mJy and with a reliable
redshift estimate, yielding a total of N$_{TOT}=52$ sources.
Figure~\ref{fig:N_zp} shows the redshift distribution of this sample
for $z\le 0.8$. The solid line is for all the objects, while the
dashed line is the distribution obtained by imposing the further
constraint R $\le 18.6$, the limiting magnitude for the completeness
of our sample.

The $N(z)$ distribution from the Phoenix sample is dominated by two
peaks at $z=0.15$ and $z=0.4$, presumably due to two major galaxy
concentrations. These features, together with the exceptional
field-to-field variations which occur in the WHT sample show how
seriously large-scale structures affect deep radio surveys. An $N(z)$
determination requires as many independent areas as possible; and to
this end we have no hesitation in joining our sample with the Phoenix
sample to make such a determination.

\subsection{Incompleteness}

We must consider the factors affecting the redshift completeness of the
samples before comparing to model predictions.

For the Phoenix survey the completeness down to 0.4 mJy is $\sim 80
\%$ in the shallower region, so we expect that it is essentially
complete to $\sim 1 $ mJy.  Thus incompleteness in the spectroscopic
sample comes only from the limiting magnitude R=22.5 which restricts
identification and spectroscopy to 47 per cent of the sample.

\begin{figure} 
\vspace{8cm}  
\includegraphics{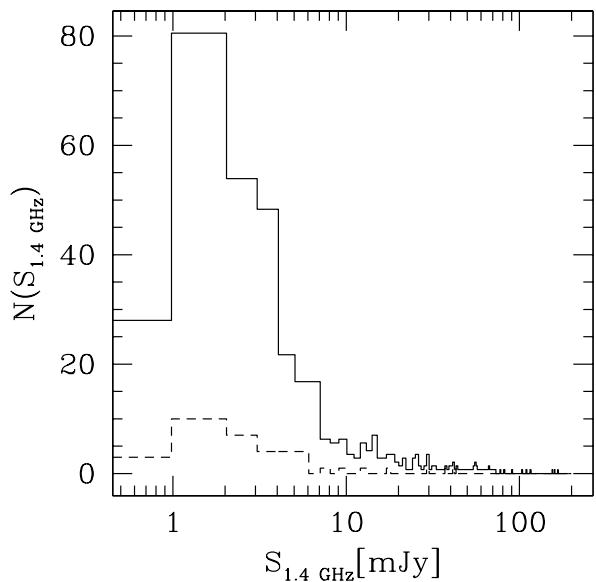}
\caption{Distribution of objects per radio flux-density S$_{1.4 \rm{GHz}}$
  interval. The solid line represents the whole spectroscopic sample,
  the dotted line those objects with a redshift determination.
\label{fig:hist_flux}} 
\end{figure}

The WHT sample has three types of incompleteness:
\begin{itemize}
\item {\bf Incompleteness in the radio catalogue.}
Becker, White \& Helfand (1995) estimated the catalogue obtained from
the FIRST survey to be 80~per~cent complete down to a flux density of
1~mJy. The issue is how this incompleteness affects the population mix
in that compact sources at 1~mJy will be present, while resolution
will preferentially lose sources of extended structure. Paradoxically
this is more of a problem for AGN and as these have been excised from
our sample by virtue of their high redshifts, this issue may be of no
significance. Starburst galaxies have relatively compact structures
(Richards, 1999) so that the resolution effects of a 5~arcsec beam
will not be serious.

\item {\bf Incompleteness in the acquisition of the spectroscopic data.}
Not all the catalogued radio-source positions in the 8 fields observed
had a fibre placed on them (Section~3 and Figures~\ref{fig:fields} and 4).
This was due (1) the geometry of the WYFFOS spectrograph, and (2) the
issue of multi-component objects (see Magliocchetti et al., 1998)
affecting $\sim 10$~per~cent of the whole sample. In these
(recognised) cases the fibre was placed at the mid-point of the
double/triple source.

\item{\bf Incompleteness in redshift determination.}
Several poorly-defined factors are involved. For instance, as the
sample was selected on the basis of radio flux alone, the process of redshift
determination is biased 
against intrinsically dusty sources; but this same bias applies to the
Phoenix sample. Intergalactic dust screening could also be involved;
see Section~6.3 regarding field-to field variation.

Figure \ref{fig:hist_flux} compares the number of sources in the whole
spectroscopic sample per radio flux interval with the relative
distribution obtained for the sub-sample with measured redshifts 
(hereafter, the `$z$-sample'). The ratio $N(S)(z)/N(S)\simeq 13$~per~cent
between the samples stays approximately constant with S, at least up
to S $\sim 15$~mJy, at which point the number of objects in the
samples becomes too small to make a meaningful comparison.

The constancy of the $N(S)(z)/N(S)$ ratio in Figure 12 implies that
there is no radio flux-bias in the process of $z$ determination. Thus
a reasonable supposition is that incompleteness in the radio survey
(resolution in particular) does not seriously affect the population mix of the
$z$-sample. We therefore suggest that apart from the known
incompleteness due to fibre coverage, the optical flux is the prime
cause of incompleteness. Though we have not explicitly imposed a
magnitude limit, the redshifts require adequate signal-to-noise in the
optical spectra. For non-emission line galaxies, this effectively
means a completeness in the continuum magnitude limit, which, as seen in
Figure~\ref{fig:mags_flux}, corresponds to R $\simeq 18.6$~mag. Using
the R-$z$ relationship of Figure~\ref{fig:mags_flux} and equation (1),
for a value of R $\simeq 18.6$, the $z$-sample (which only includes 
objects with radio fluxes $S_{\rm 1.4 GHz}\ge 1$ mJy) is complete in redshift
up to $z\simeq 0.3\pm 0.1$. The uncertainty corresponds to the scatter
in the R-$z$ relation. The procedure cannot be applied to Seyfert1
objects, which do not follow the $R-z$ relation. However our
conclusions are not affected given that this class of objects is
mainly found at redshifts $z>0.3$.

As a further test, Figure~\ref{fig:N_zp} shows the variation in the
$N(z)$ for the Phoenix survey if we consider a magnitude cut of
R=18.6 (dashed line). The number of sources per redshift is
unaffected to $z=0.25$. The percentage then starts dropping as the
R=18.6 magnitude limit comes into play. From a similar analysis using
our R-$z$ relation, we find that the Phoenix survey sample is
essentially complete to $z\sim0.9$ for objects other than radio AGN.

\end{itemize}

\subsection{Comparison with models and results}

The $N(z)$ model predictions used here are from the study by Dunlop \&
Peacock (1990; hereafter DP) in which Maximum Entropy analysis was
used to determine the coefficients of polynomial expansions
representing the epoch-dependent radio luminosity functions of radio
AGNs. The analysis incorporated the then-available identification and
redshift data for complete samples from radio surveys at several
frequencies. Indeed the 7 derivations of luminosity functions carried
out by DP predict source counts and $N(z)$ distributions for
frequencies 150~MHz to 5~GHz down to 100~mJy with considerable
accuracy, the input data spanning approximately this parameter
space. However, extrapolating down to mJy levels, the models show
large variations in the predicted $N(z)$; see Figure 1 of
Magliocchetti et al., 1999. Nevertheless, all predictions show a broad
peak with median redshift of about 1.  Some models also produce a
'spike' at small redshifts indicating that at such low flux densities,
the lowest-power tail of the local radio luminosity function begins to
contribute substantial numbers of low-redshift sources. Note that the
DP formulation did not encompass any evolving starburst-galaxy
population explicitly; it was restricted to two radio AGN populations,
the 'steep-spectrum' extended-structure (FRI and FRII) objects and the
'flat spectrum', compact objects (predominantly Seyfert1/quasars and
BL\,Lacs). Hence, at some level, the low-$z$ spike must be considered
as a fortunate artifact of the models.

To compare the predictions from DP models with the data we have to
consider the incompleteness effects described above.  With regard to
normalization of surface density, the DP predictions give $508\la
N_{TOT}({\rm S}\ge 1\ \rm{mJy})\la 570$ for an area of $8\times(\pi\;0.5^2)$
square degrees. For our 8 chosen fields of the FIRST survey, we found
N=525 (after correcting for the presence of multicomponent sources),
in excellent agreement.
\begin{figure} 
\vspace{8cm}  
\includegraphics{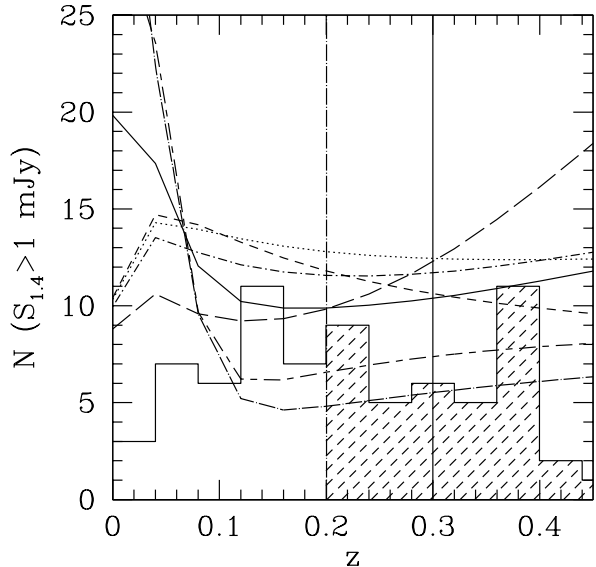}
\caption{Distribution of objects as a function of redshift for the
  WHT + Phoenix sample. The shaded area corresponds to the redshift
  range $z\simeq0.3\pm 0.1$ 
where the completeness of the sample is less than $100\%$
(see text for details). Curves show the different DP models:
model 1, dotted line; model 2, short dashed line; model 3,  
long dashed line; model 4, dot-short dashed line; 
model 6, dot-long dashed line; model 7,
short dashed-long dashed line; average, solid line. Model 5 is omitted 
because it has an imposed redshift cutoff at z=4 which results in an 
unrealistic sharp and dominant feature at this redshift.
\label{fig:N_zgt1}} 
\end{figure}

Figure \ref{fig:N_zgt1} shows the distribution of objects in the
WHT+Phoenix sample (76 objects up to $z= 0.52$) as a function of
redshift. The shaded area corresponds to the redshift range $z\simeq
0.3\pm 0.1$ where the combined sample starts losing completeness due
to magnitude incompleteness in the WHT sample. The smooth curves are
the predictions from DP models. None of these provides a good fit. The
percentage of objects at $z\la 0.1$ is seriously overestimated in all
the models, especially 6 and 7 that feature an unrealistic $z\sim 0$
spike, presumably an artifact caused by extrapolating too far. Though
the uncertainties are large, only one model (model 3) roughly follows
the steady rise in $N(z)$ to $z\simeq 0.2$, while all the others
either show a plateau or decrease for $z\ga 0.1$. Note that model 3 is
also the closest to the correct low-$z$ normalisation.

Our data show that the population of starburst galaxies constitutes
only a small fraction of the radio objects at the mJy level, contrary
to early claims (e.g. Windhorst et al., 1985).  If the majority of
objects at radio fluxes 1~mJy~$\la S\la 3$~mJy were starburst galaxies
we would have obtained redshifts for all of them. In fact such objects
are on average the brightest in apparent magnitude and the closest in
the $z$-sample (see also Benn et al., 1993; Gruppioni, Mignoli \&
Zamorani, 1998; Georgakakis et al., 1999). 
%
%
%
Allowing for the sensitivity limit of our redshift determinations, and
the strong emission line spectrum for this type of objects we would expect 
$\sim$100~per~cent redshift completeness up to $z\simeq 0.2$.
If indeed they constituted most of the sources at 1~mJy, we would have
therefore found a ratio $N(S)(z)/N(S)$ to be $\sim 1$ at low radio fluxes. 
Figure \ref{fig:hist_flux} shows that this is not the case; all flux
intervals have the similar probability ($\sim 13$~per~cent) of
redshift determination. Thus only $\sim 5$~per~cent of the population
at S$_{1.4\,{\rm GHz}}$ = 1~mJy can be starburst objects; the great
majority of identifications remain AGN, with predominantly early-type
galaxies as hosts, too faint in optical magnitude to be detected in
our spectroscopic observations. Very similar conclusions are reached
by Gruppioni, Mignoli \& Zamorani, 1998 and Georgakakis et al., 1999
for their mJy/sub-mJy samples, in which the optical observations were
pushed down to R=24 and R=22.5 magnitudes respectively.

We stress again that we did not impose a magnitude limit before our
spectroscopic observations, so that any optically-faint star-forming
galaxies would have been observed in our sample.  The presence of
emission lines would have allowed us to get redshifts for those
sources much fainter than the APM survey magnitude limit. However, we
found that that for R $\ga 17$ the only classes of objects
detected were early-type and Seyfert1 galaxies. This makes the
constraint on the percentage of star-bursting galaxies at mJy level
yet more robust.

We note that our determination of $N(z)$ at low $z$ provides
an additional datum on the shape of the complete $N(z)$
distribution. The overall normalisations between our sample and the DP
predictions agree; yet the DP predictions significantly overestimate
$N(z)$ at $z< 0.4$. This disagreement at low redshifts may imply that we
expect to find many more sources at $z\ga 0.3$, perhaps consistent
with the DP model 3. 

However there is a significant uncertainty caused by the large
variation in fraction of optically identified sources in different 1
degree WHT fields (the fraction varies by a factor 4, see
Table~\ref{Table_fields}). The combined Phoenix and WHT samples are
equivalent to roughly 10 times the area of one WHT field, so assuming
Poisson statistics we expect roughly a factor 1.5 uncertainty in the
overall surface density. Allowing for such variations, any of the DP
models 1-4 too are reasonably consistent with the data.

\section{CONCLUSIONS}

We have carried out multi-object spectroscopy of an unbiased selection
of FIRST radio sources (S $\ga 0.8$~mJy) by placing fibres at the
positions of 365 sources ($\sim$ 69~per~cent of the complete radio
sample). The spectra obtained have enabled us to measure 46 redshifts,
$\sim 13$~per~cent of the targeted objects.  APM data have provided 
morphology and photometric data for the corresponding optical
identifications. The photometry shows that redshift measurements were
obtained only for objects brighter than R $\simeq 20.5$~mag; from the
tight R-$z$ relation observed, the redshift sample is estimated to be
$\sim$100~per~cent complete to R=18.6~mag.

The objects in the spectroscopic sample with R $\la 20.5$ are a mixture
of early-type galaxies at relatively high redshifts, $z\ga 0.2$ ($\sim
45$~per~cent of the sample), late-type galaxies at intermediate
redshifts, $0.02\la z\la 0.2$ ($\sim15$~per~cent), and very local
starburst galaxies with $z\la 0.05$ ($\sim 6$~per~cent). We also found
a number of Seyfert1/quasar type galaxies, all at $z\ga 0.8$ ($\sim
9$~per~cent of the sample), two Seyfert2's (4~per~cent), and 4
stars. The number of objects with featureless spectra are most
probably early-type galaxies, given the shape of the continuum, the
lack of emission lines and the red colours. Using the R-$z$ relation
derived for our sample, we conclude that they are likely to have $z\ga
0.3$.
The redshift incompleteness does not depend on radio flux density;
optical apparent magnitude is the only identifiable factor. Using
again the R-$z$ relation determined for the sample, we estimate
$\sim$ 100~per~cent completeness for the spectroscopic sample up to
$z\sim0.3\pm 0.1$.

To define $N(z)$ at S$_{1.4\,\rm{GHz}}$ as well as possible, we have
combined our sample with the Phoenix spectroscopic sample
(Georgakakis et al., 1999), which we estimate to be complete (for
non-AGN objects) to $z=0.9$. The combined distribution
(Figure~\ref{fig:N_zgt1}) shows the following:

\begin{itemize}

\item 
The redshift distribution rises up to $z\simeq 0.05$
and is then approximately leveled to $z=0.3$.

\item 
The total number of sources predicted by the luminosity-function
models of Dunlop \& Peacock (1990) agrees with that observed.

\item 
None of the models provides a good fit to the shape of $N(z)$.
The percentage of objects at $z\la 0.1$ is seriously overestimated in
almost all the models, especially for the pure-luminosity and
luminosity/density evolution models (DP 6 and 7) that feature an
unrealistic $z\sim 0$ spike.

\item 
The normalization of the models appears to be too high to fit the
observed $N(z)$ for $z \la 0.2$.  This disagreement may imply that the
model shape is wrong, and there are more sources at $z\ga 0.3$ than
indicated by the models.  Alternatively, the discrepancy could be due
to observing a low density by chance, given a large variance in galaxy
density caused by strong clustering.

\item
The $N(z)$ at $S_{1.4\,{\rm GHz}} = 1$~mJy is dominated by AGN, and
starburst objects constitute less than 5~per~cent of the total. This
is a robust conclusion. More starburst galaxies would have
substantially raised the proportion of objects with redshift
determinations; and a significant intrusion of starburst galaxies at
the lowest radio flux densities would have resulted in a higher
success rate in redshift determination with decreasing flux
density. The great majority of objects in the sample at this level are
AGN associated with early-type galaxies whose optical continua and
weak-to non-existent emission lines place them at or below the limit
of our spectroscopic survey.

\end{itemize}

The accurate definition of the low$-z$ end of the $N(z)$ relation has
impact in four areas: (i) the population mix, which is critical for
testing and refining dual-population unified models; (ii) the
definition of the local luminosity functions, important for modelling
both the form and cosmic evolution of the overall luminosity
functions; (iii) the derivation of spatial measurements of the
large-scale structure from angular measurements; and (iv) constraints
which it enables to be placed on the global star-formation-rate up to
$z=0.3$. 

\vspace{1cm}
\noindent
{\bf ACKNOWLEDGEMENTS}\\ 
MM acknowledges support from the Isaac Newton Scholarship. We thank
David Helfand for extremely helpful discussions.  The WHT is operated
on the island of La Palma by the Isaac Newton Group in the Spanish
Observatorio del Roque de los Muchachos of the Instituto de
Astrofisica de Canarias. GC acknowledges a PPARC Postdoctoral Research Fellowship.

\end{document}